\documentclass[11pt,english]{article}
\usepackage[letterpaper]{geometry}
\usepackage{amsmath}
\usepackage{graphicx}
\usepackage[numbers]{natbib}
\usepackage{xcolor}

\makeatletter
\usepackage {url}
\date{}
\makeatother
\usepackage{babel}

\begin{document}

\title{Comparing optimization strategies for force field parameterization}
\author{Fatih G.~Sen$^{1}$, Badri Narayanan$^{1}$, Jeffrey Larson$^{2}$, Alper Kinaci$^{1}$, Kiran Sasikumar$^{1}$,\\
Michael J.~Davis$^{3}$,  Stefan M.~Wild$^{2}$, Stephen K.~Gray$^{1}$,\\ Subramanian K.~R.~S.~Sankaranarayanan$^{1}$, Maria K.~Y.~Chan$^{1}$}
\maketitle
$^{1}$Center for Nanoscale Materials, $^{2}$Mathematics and Computer Science Division, $^{3}$Chemical Sciences and
Engineering Division, Argonne National Laboratory, 9700 Cass Ave,
Lemont, IL 60439, USA

\subsection*{Abstract}
Classical molecular dynamics (MD) simulations enable modeling of materials and examination of microscopic details that are not accessible  experimentally. The predictive capability of MD relies on the force field (FF) used to describe interatomic interactions. FF parameters are typically determined to reproduce selected material properties computed from density functional theory (DFT) and/or measured experimentally. 
A common practice in parameterizing FFs is to use least-squares local minimization algorithms. Genetic algorithms (GAs) have also been demonstrated as a viable global optimization approach, even for complex FFs. However, an understanding of the relative effectiveness and efficiency of different optimization techniques for the determination of FF parameters is still lacking. 
In this work, we evaluate various FF parameter optimization schemes, using as example a training data set calculated from DFT for different polymorphs of Ir$O_2$. The Morse functional form is chosen for the pairwise interactions and the optimization of the parameters against the training data is carried out using (1) multi-start local optimization algorithms: Simplex, Levenberg-Marquardt, and POUNDERS, (2) single-objective GA, and (3) multi-objective GA. 
Using random search as a baseline, we compare the algorithms in terms of reaching the lowest error, and number of function evaluations. We also compare the effectiveness of different approaches for FF parameterization using a test data set with known ground truth (i.e generated from a specific Morse FF). We find that the performance of optimization approaches differs when using the Test data vs. the DFT data. Overall, this study provides insight for selecting a suitable optimization method for FF parameterization, which in turn can enable more accurate prediction of material properties and chemical phenomena.


\section{Introduction}
Classical molecular dynamics (MD) is a widely-used technique for the atomistic modeling of materials and macromolecular systems. With the exponential increase in computing power, MD simulations with relatively simple forms of force fields (FFs) can easily describe structure and dynamics in multi-million to billion-atom systems \citep{Abraham2002,Kadau2002,Vashishta2008,Shekhar2013,Berman2015}. 
The reliability of MD rests heavily on the accuracy with which the empirical FFs describe the interatomic interactions and in particular, the actual forces in the system. Since direct experimental measurements of interatomic forces are usually not available, FFs are typically parameterized to reproduce interatomic interactions obtained from quantum-mechanical methods such as density functional theory (DFT), in addition to a handful of experimentally-accessible observables. There have been tremendous efforts in the development of the mathematical form of \citep{Allen1989computer,Tersoff1988,Daw1993eam,Catlow1994,Finnis2007bop,Sinnott2012},
as well as approaches to obtain the best set of parameters for, the force fields \citep{Ercolessi1994,Gale03,Waldher2010forcefit,Artrith2014neural,Grimme2014qmdff,Jaramillo2014garffield,Thompson2014,Wang2014FB}. However, systematic evaluation of different optimization techniques for the determination of such force field parameters are lacking in literature. In this work, we test the efficiency of a variety of global and local optimization schemes in the parameterization of FFs.

Over the past decades, FFs with various functional forms have been suggested for different chemical systems. To describe interatomic interactions, there are a wide range of possibilities, from simpler functional forms such as Lennard-Jones (L-J) \citep{Jones1924}, Morse, and Buckingham \citep{Buckingham1938}, to the more complex reactive force fields (e.g. ReaxFF). For ionic systems, simple non-bonded functions such as Morse and Buckingham are often coupled with a Coulomb term to account for electrostatic interactions \citep{Catlow1994}. For metallic, covalent, and multi-component systems, many body interactions were included in Finnis-Sinclair \citep{Finnis1984}, Stillinger-Weber \citep{Stillinger1985}, embedded-atom method (EAM) \citep{Daw1993eam}, modified-embedded atom method (MEAM) \citep{Baskes1992meam}, Tersoff \citep{Tersoff1988}, and bond-order \citep{Brenner1990bop,Brenner2002rebo} FFs. 
Recent developments, which combine bond-order concepts with charge transfer schemes that predict local chemical environment in reactive simulations, have allowed for a description of reactive interfaces within MD framework. The two most common forms of such reactive force fields are: i) ReaxFF (reactive force field) method developed by van Duin and coworkers \citep{vanDuin2001reaxff,Chenoweth2008reaxff}, and (ii) charge-optimized many-body (COMB) FFs developed by Sinnott, Phillpot, and coworkers \citep{Shan2010comb,Devine2011comb}. 
Depending on the complexity of the functional form and the number of species involved in the system, the number of parameters that describe the interatomic interactions in an FF varies from only two parameters per pair in an L-J to more than hundreds of parameters in ReaxFF. Accordingly, determination of force-field parameters requires an efficient and systematic approach, which can sample the given parameter space effectively.

In the past, many different optimization algorithms have been used to parameterize FFs including local and global optimization methods. COMB \citep{Shan2010comb} and ForceFit \citep{Waldher2010forcefit} package implemented least-squares methods \citep{Powell1965}, GULP \citep{Gale03} implemented Simplex \citep{Nelder1965} and BFGS \citep{Broyden1970,Fletcher1970,Goldfarb1970,Shanno1970} methods and ForceBalance \citep{Wang2014FB} implemented Levenberg-Marquardt
(L-M) \citep{More1978LM} method to fit FF parameters. ReaxFF \citep{vanDuin2001reaxff} is reported to be developed with a one-parameter search method \citep{vanDuin1994}. 
Although local optimization methods have been shown to be successful in parameter fitting, they rely greatly on the initial guess for the unknown parameters and can fail for very complex functional forms. To cover a significant volume of the parameter phase space, large number of starting guesses need to
be evaluated
. On the other hand, global optimization methods intrinsically search the unknown parameter phase space and can be more suitable for FF parameterization. Genetic algorithms (GA) can offer an unbiased global search by maintaining a diverse population, hence discovering potentially good regions in the parameter space \citep{Diwekar2008opt}. Owing to the probabilistic development of the solution, it should be noted that GA also does not guarantee optimality. 
In addition to genetic algorithms (GA) \citep{Kinaci2012,Saito2001,Narayanan2015au,Cherukara2016,Sen15}, other global optimization methods are also employed for parameterizing FFs including particle swarm optimization (PSO) \citep{Cui2012pso,Kandemir2016pso}, neural networks \citep{Hobday1999,Artrith2014neural}, 
and database optimization \citep{Zhang2015}.

In all local and global optimization schemes, the goodness of fit, and hence the predictive capability of the force field, depends on the weighting scheme selected between different observables (i.e. properties to be reproduced by the FF). In selection of weights, one may consider the difference in the magnitudes of observables, number of different categories of observables (such as energies, elastic constants, lattice parameters, etc.) and number of quantities within a categories of observable as well as the order in relative importance of of the observable for the problem of interest.
One way of removing the uncertainties due to weighting is to use a multi-objective optimization algorithm \citep{Diwekar2008opt}. In multi-objective optimization, if no objective can be improved without sacrificing at least one other, a set of non-dominated solutions are obtained, which is called the Pareto optimal solution set. Recently, a multi-objective genetic algorithm (MOGA) method has been implemented in GARFFIELD \citep{Jaramillo2014garffield} package to parameterize reactive force fields. Such MOGA approaches are useful in selection of force field parameters favoring a specific property without having to re-optimize for that property with an appropriate weighting \citep{Narayanan2016zrn}. Furthermore, it is easier to identify parameter-property relationships.

Overall, many different optimization methods have been used in earlier studies to develop FFs with successful results. 
However, since the focus in almost all prior studies on FF development is limited to one particular optimization method of choice, one could not infer the relative efficiency and accuracy of different optimization algorithms. In the present study, we fill in this gap by systematically comparing different local and global optimization methods in parameterizing a model FF based on pairwise Morse interactions to describe various crystal polymorphs of IrO$_2$. IrO$_2$ is an exciting electrocatalyst for water splitting reaction to produce hydrogen from water using sunlight \citep{Comninellis1991,Beni1979,Song2008,Suntivich2011,lee2012synthesis}. A previous study demonstrated that Morse FF coupled with a variable charge (QEq) scheme can accurately describe bulk, surface and catalytic properties of IrO$_{2}$. Here, we will employ only the Morse FF to simplify the problem for testing purposes. We also investigate benefits of using multi-objective algorithm to remove the ambiguities brought by weighting, and to tailor FFs to particular properties. To compare the effectiveness of different algorithms when the ground truth is known vs. when it is not, we compare the algorithms for parameter optimization when using a test training set from pre-specified Morse FF parameters vs. when using a DFT-generated training set. Finally, we provide suggestions for an efficient optimization method for force field fitting.

\section{Methodology}
\subsection{Training dataset generation}
The aim in force field parameterization is to find a parameter set that can reproduce measured or calculated values of specific properties of the system. Moreover, the developed force field should also be able predict properties that were not in the training set. Hence, the training set for fitting should be extensive and consist of many different properties, which will increase the transferability of the FF. 
In the present work, as a first step towards optimization, an extensive training data set is generated from structural, thermodynamic and mechanical properties of various IrO$_{2}$ crystal polymorphs that are computed using density functional theory (DFT). The DFT training set includes unit cell energies, lattice constants, internal coordinates and elastic constants of rutile ($P4_{2}/mnm$), pyrite ($Pa-3$), anatase ($I4_{1}/amd$), columbite ($Pbcn$) and brookite ($Pbca$) phases of IrO$_{2}$. In total, the training set consists of $216$ energies, $8$ elastic constants, $9$ lattice constants, and $15$ internal coordinates.  

All DFT calculations are performed using the projector-augmented wave (PAW) method \citep{Blochl94}, as implemented in the Vienna Ab-initio Simulation Package (VASP) \citep{Kresse94,Kresse96}, and with the generalized gradient approximation (GGA) of the exchange correlation described by Perdew, Burke, and Ernzerhof (PBE) \citep{Perdew96}. 
Calculations are carried out using PAW\textendash PBE atomic potentials supplied with the VASP code \citep{Kresse99,Blochl94}, and include the Hubbard $U$ correction \citep{Liechtenstein95,Dudarev98}, which is beneficial for describing unfilled $d$-shells. Spin-orbit coupling (SOC) is also included since it has significant impact on the properties of Ir \citep{Fujiwara13,Panda14}.
All atomic positions and unit cell parameters are relaxed using $10\times10\times10$ $\Gamma$-centered k-point grids, and a kinetic energy cutoff of $520$
eV. These $k$-point and energy cutoff settings ensure total energy convergence to $1-2$ meV/atom. We use a Hubbard $U$ parameter of $1.0$ eV, which was found to give accurate ground state properties and formation enthalpies \citep{Sen15}. 

In addition to the DFT-based training set, we calculate the same observables using a specific Morse parameter set. This training set, referred to as "test data" in the following text, has a known ground truth (i.e. solution with absolute zero error), and we use it to compare efficiency and performance of different optimization algorithms in cases where ground truth is known vs. where it is not.  This is of importance because the performance of optimization approaches may differ in the two situations, due to systematic as well as random errors in the data when computed with DFT. 

\subsection{Optimization approaches}
\begin{figure}
\centering{}\includegraphics[width=10cm]{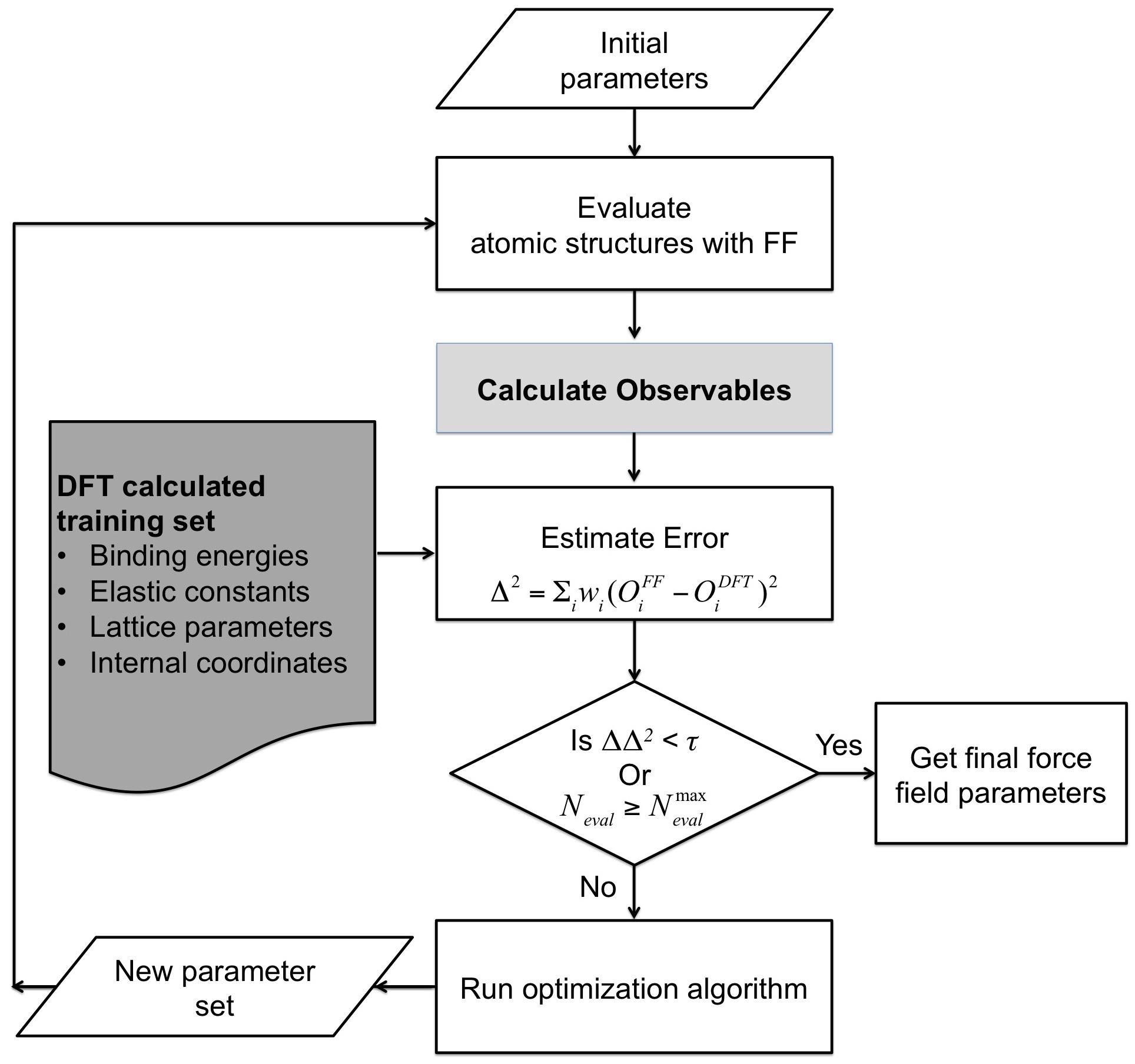}\protect\caption{Flowchart for the parameterization of force fields.\label{fig:flowchart}}
\end{figure}

A first-principles based force field parameterization process involves adjusting the variables of the FF using optimization techniques to minimize the error in reproducing the desired observables in the training set. In Figure \ref{fig:flowchart}, a flowchart for the optimization process is shown. Optimization starts with an initial parameter set, which is randomly determined. For each set of parameters, we evaluate properties of solid IrO$_{2}$ for each atomic structure using a simple sum of two-body interactions,

\begin{equation}
V( \mathbf{r}_1, \mathbf{r}_2, ..., \mathbf{r}_M ) 
= \sum_p^M \sum_{q > p}^{M-1} v_{pq} ( r_{pq} )  ~~~,
\end{equation}

\noindent where $M$ is the number of atoms in the solid structure. We take the pairwise potentials $v_{pq}$ to be Morse functions of the interatomic separation $(r_{pq})$ between atoms $p$ and $q$ as implemented in General Utility Lattice Program (GULP) \citep{Gale03},

\begin{equation}
v_{pq}(r_{pq})=D_e^{pq}\left(\left[1-\mbox{exp}(-\alpha^{pq}(r_{pq}-r_e^{pq}))\right]^{2}-1\right)\label{eq:morse}
\end{equation}

\noindent where $D_{e}^{pq}$ is the bond dissociation energy, $\alpha^{pq}$ is related to the force constant $k^{pq}$ via $\alpha^{pq} = \sqrt{k^{pq}/(2D_e^{pq})}$, and $r_e^{pq}$ is the equilibrium interatomic distance, all associated with the $p,q$ atomic pair. In a binary system such as
Ir-O, there are three pair types, namely Ir-Ir, Ir-O, and O-O. Therefore, in this work we have 3 Morse parameters for each of the 3 atomic pair types, and in
total there are 9 parameters to determine.

Once the observables (${O_{i}^{FF}}$) are evaluated based on the FF with parameter set, they are compared with the corresponding DFT values (${O_{i}^{DFT}}$) to determine the weighted sum of squares of errors ($\Delta^{2}$), with weights ${w_{i}}$. Then, the optimization algorithm is called, which generates a new set of parameters to evaluate the properties, and reduces $\Delta^{2}$. Ideally, ${w_{i}}$ values should be set such that each category of the properties (energies, elastic constants, lattice constants and internal coordinates) are weighted approximately equally in the total sum of squares of errors. However, the size of the errors in each property is not known \textit{a priori}. We determine${w_{i}}$, using as proxy for $\Delta_n^2 \equiv \sum_{i\in n} w_i (O_i^{FF} - O_i^{DFT})^2$ the squared errors (compared to experimental measurements) reported for different types ${n}$ of properties in DFT calculations, from requiring that $\Delta_{\text{Energy}^2} = \Delta_{\text{Elastic}^2} = \Delta_{\text{Lattice}^2} = \Delta_{\text{Internal}^2}$. 
From the available literature, the errors in DFT values are estimated for energies, elastic constants, lattice constants, and internal coordinates as $\Delta_{\text{Energy}}=0.44$ eV/IrO$_{2}$ \citep{Jain13MatProject}, $\Delta_{\text{Elastic}}=14.2$ GPa \citep{Mattsson08AM05}, $\Delta_{\text{Lattice}}=0.046$ Å \citep{Mattsson08AM05,Heyd04HSE}, $\Delta_{\text{Internal}}=0.004$, respectively. Accordingly, we obtain normalized weights for different properties as 
$w_{\text{Energy}}=1.0$, $w_{\text{Elastic}}=2.5\times10^{-2}$, $w_{\text{Lattice}}=2.2\times10^{3}$ and $w_{\text{Internal}}=1.7\times10^{5}$. We note that there the discrepancies between DFT and experiments are not necessarily related to the discrepancies between DFT and FF evaluations, so the above scheme is far from ideal. The lack of a prior estimate of ${w_i}$, however, also illustrates the problematic nature of weighting errors of different categories in FF parameterization. 

\subsection{Optimization algorithms}
We consider three local optimization algorithms, Simplex, Levenberg-Marquardt (L-M), and Practical Optimization Using No Derivatives for sums of Squares (POUNDERS). 
Simplex \citep{Nelder1965,Press2007} is a commonly used method for function minimization, in which a simplex with  $(N+1)$ vertices is constructed for a function with $N$ variables. At each iteration, a vertex with the highest value is replaced with another point. During iterations, simplex contracts on to the final
minimum. Simplex algorithm requires only function evaluations, and does not require function derivatives. Therefore, it is known to require more function evaluations than the methods that use derivatives such as Levenberg-Marquardt (L-M)\citep{Huang1998Simplex}. 
L-M \citep{More1978LM} is also a common optimization algorithm for non-linear least squares fitting problems. In L-M, a linear Taylor expansion
approximation is used to estimate value of the function at a given point. An iterative process is carried out to minimize the difference in function values between two consecutive iterations. We used L-M algorithm as implemented in MINPACK \citep{More1980minpack,More1984minpack}.
Practical Optimization Using No Derivatives for sums of Squares (POUNDERS) is an algorithm that uses the availability of the vector of residuals
$w_i(O^{FF}_i - O^{DFT}_i)$ rather than just the weighted sum-of-squares error $\Delta^2$ to build models of each residual as functions of the 9 optimization
parameters. POUNDERS then generates candidate points that minimize a a combination of the residual models; the combination follows the known least-squares form of  the problem. By using this information, POUNDERS is often able to more quickly optimize least-squares problems than do methods that only work with the combined  residual. For the mathematical background of POUNDERS, see \cite{SWCHAP14}; for an application of POUNDERS to a parameter estimation problem in nuclear density  functional theory, see \cite{WSS15}. The POUNDERS algorithm is included in the distribution of PETSc/TAO, an open-source software package \cite{Petsc}.

As a global optimization method, we test a simple genetic algorithm (SGA) for comparison to the local search algorithms. SGA is a heuristic method that mimics natural evolution. GA is based on the survival of the best individuals at every generation. Here, every parameter set is considered as an individual, which is a part of a population. We perform genetic operations on the population, namely tournament selection without replacement \citep{Goldberg1989,Sastry01}, simulated binary crossover \citep{Deb95-crossover,Deb95-crossover2}, and polynomial type mutation of order 20 \citep{Deb95-crossover,Deb95-crossover2}, to generate a new population of parameter sets. SGA optimization was carried out using the GA Toolbox\citep{Sastry07GA} with a crossover probability of $0.9$ and mutation probability of $0.1$. As an additional variant, Simplex is performed during GA for further optimization.  

We also test a multi-objective genetic algorithm (MOGA) to remove ambiguities caused by the selection of weights in the single objective optimization. We divide the optimization problem into 4 objectives and group the sum of squares of errors for each type of property as a separate objective (energy, elastic constants,
lattice constants and internal coordinates), and within each objective we use equal weighting of $w=1.0$. We utilize a non-dominated sorting algorithm (NSGA-II) \citep{Deb02-nsga2,Sastry07GA}, which dictates that a candidate individual is kept in the population through successive generations, if it is non-dominated, i.e. there exists no other candidate that is better in all objectives. As the generations progress, a Pareto front of non-dominated candidate solutions form. Eventually, when the Pareto front ceases to advance, one can select the desired solution among the solutions set on Pareto front. The selection of a single solution from the Pareto optimal set can be carried out using weighted sum method, goal programming, constrain methods or clustering algorithms \citep{Diwekar2008opt,Ehrgott2005opt,Konak2006multi}.

As a baseline for comparisons, we evaluate 500000 random set of parameters. 
For Simplex, L-M, and POUNDERS, we employed a multi-start scheme to obtain a global optimum and generated $500$ initial  guess sets of parameters that are
selected randomly within the space given by the ranges for parameters and used the same $500$ parameters for all Simplex, L-M, and POUNDERS methods.
For SGA, we consider different population sizes of $100$, $400$ and $1000$. 
The ranges for the parameter search space used in both local and SGA methods are $D_{e}^{Ir-Ir}:0.001-0.01$ eV, $\alpha^{Ir-Ir}:1.0-3.0$ \AA$^{-1}$, $r_{0}^{Ir-Ir}:4.0-6.0$ \AA, $D_{e}^{Ir-O}:1.5-3.2$ eV, $\alpha^{Ir-O}:1.7-3.0$ \AA$^{-1}$, $r_{0}^{Ir-O}:1.8-2.0$ \AA, $D_{e}^{O-O}:0.01-0.1$ eV, $\alpha^{O-O}:1.0-2.0$ \AA$^{-1}$, $r_{0}^{O-O}:3.2-4.0$ \AA. 

In Figure \ref{fig:flowchart}, the optimization process proceeds until one of the two criteria is satisfied: i) the change in $\Delta^{2}$ between two consecutive iterations is less than a preset tolerance $(\tau)$, or ii) the number of objective function evaluations $(N_{eval})$ is greater than the maximum number of evaluations ($N_{eval}^{max}$).
For Simplex and L-M, all calculations reach the stopping criterion of $\tau<10^{-14}$ before the $N_{eval}^{max}=10000$ evaluations. For POUNDERS, we set  $N_{eval}^{max}=2000$ and used a positive trust region radius of $0.1$. We set $N_{eval}^{max}=50000$ and $100000$ as the stopping criteria for single objective and multi-objective GA respectively.

\section{Results and discussion}
\subsection{Multi-start Local Optimization Results}
\subsubsection{Test Data}
In most cases, the primary concern is the effectiveness of the optimization algorithm in attaining a parameter set with the lowest error compared to the training set, i.e. $\Delta^{2}$. For the same $500$ random initial guesses, using the Test data, the lowest $\Delta^{2}$ obtained for Simplex, L-M, and POUNDERS are $2.97$, $0.83$, $0.003$, respectively, which are very low compared to the lowest $\Delta^{2}=1140.63$ obtained with random search of $500000$ parameters. The total number of function evaluations, $N_{eval}$ for Simplex, L-M, and POUNDERS are $570020$, $105251$, and $1075000$, respectively. Accordingly, POUNDERS achieves the lowest $\Delta^{2}$, which is very close to the solution, but requires the maximum number of function evaluations among the three methods. 

\begin{figure}
\begin{centering}
\includegraphics[width=12cm]{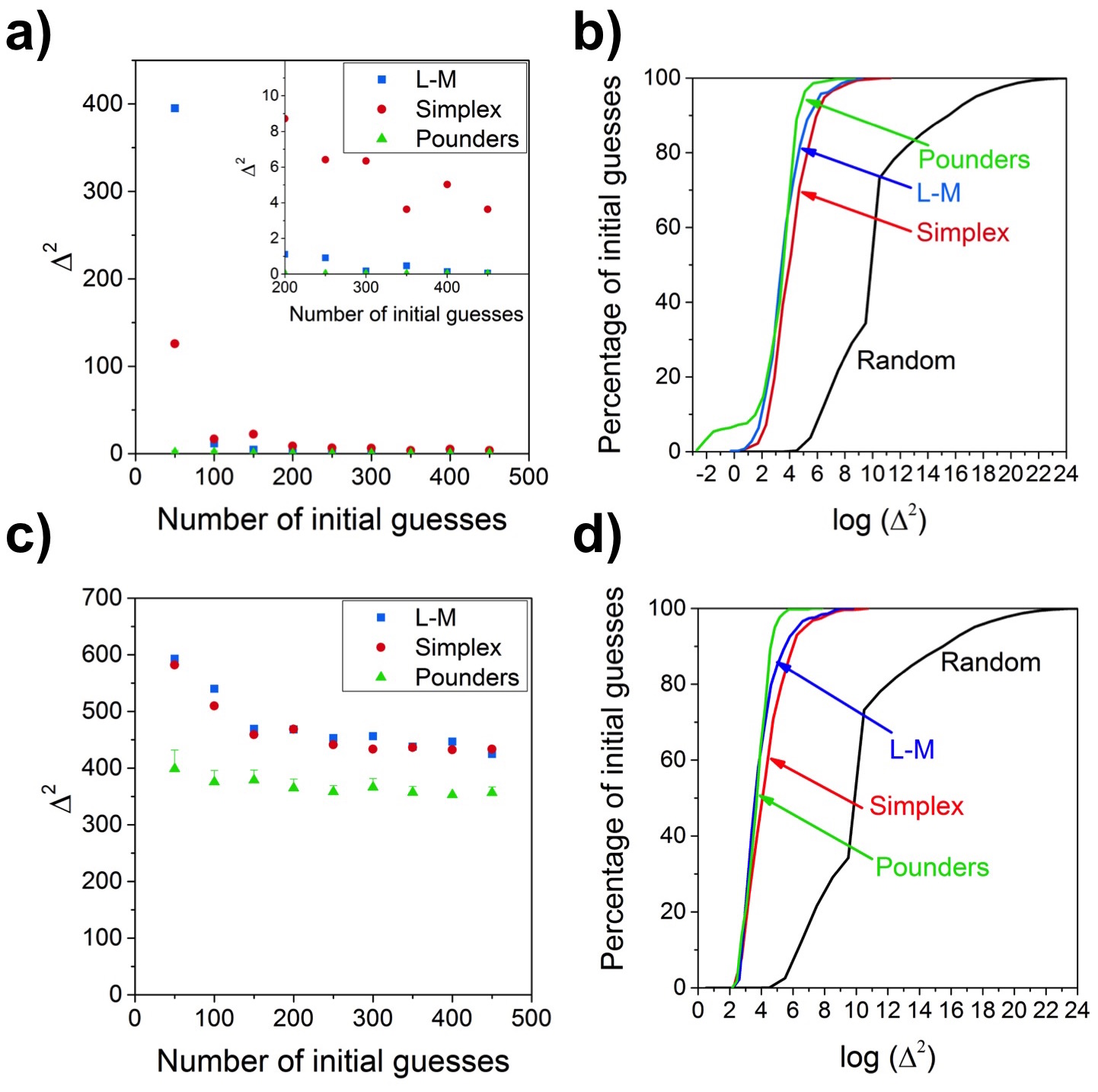}
\par\end{centering}
\protect\caption{Comparison between Simplex, Levenberg-Marquardt (L-M), and POUNDERS algorithms. For test data: a) the change in the sum of squares of errors ($\Delta^{2}$) obtained with different number of initial guess, and b) statistical count analysis of $\Delta^{2}$ values with respect to percentage of initial guesses and compared to random search. For DFT data, the same analyses are given in (c)-(d). \label{fig:lm-vs-simplex}}
\end{figure}

It is also important to determine how many initial guesses are necessary to obtain a good FF parameter set. To find the relationship between the number of initial guesses and the minimized $\Delta^{2}$ values, from the total $500$ initial guesses, we record minimized $\Delta^{2}$ values at randomly-selected intervals of $50$ up to $450$. We repeat the selection $10$ times at each interval and record the lowest $\Delta^{2}$ among the selection. Then, the average $\Delta^{2}$ of the $10$ is identified as the lowest $\Delta^{2}$ for the corresponding number of initial guesses.
The change in the $\Delta^{2}$ with respect to number of initial guesses for Simplex, L-M, and POUNDERS are shown as in Figure \ref{fig:lm-vs-simplex}(a). Figure \ref{fig:lm-vs-simplex}(a) indicates that for each number of guesses, POUNDERS results in the lowest $\Delta^{2}$. For small number of initial guesses less than 100, Simplex results in lower $\Delta^{2}$ values than L-M, but if larger number of guesses are used, L-M reached lower $\Delta^{2}$ values than Simplex. If more than 300 guesses are employed, POUNDERS and L-M reached similarly low $\Delta^{2}$ value.

Statistical count analysis of $\Delta^{2}$ values from Simplex, L-M, and POUNDERS with respect to number of initial guesses and comparison with random search
of parameters containing $500000$ sets are shown in Figure \ref{fig:lm-vs-simplex}(b). Figure \ref{fig:lm-vs-simplex}(b) indicates that Simplex, L-M, and POUNDERS can reduce the $\Delta^{2}$ by $4-8$ orders of magnitude compared to random search, indicating that all methods are greatly helpful towards achieving a good solution. POUNDERS is shown to be the most effective and efficient, as it is found to achieve a final solution with a $\Delta^2$ more than $2$ orders of magnitude lower than those from the other two approaches, from approximately $10\%$ of the initial guesses.

\subsubsection{DFT data}
If DFT data is used to parameterize the force field for the same $500$ random initial guesses, the lowest $\Delta^{2}$ obtained for Simplex, L-M, and POUNDERS are obtained as $431.89$, $418.59$, $352.79$, respectively, which are substantially lower than the lowest $\Delta^{2}=27722.03$ found with random search of $500000$ parameter sets. Since there is no known solution for the DFT dataset, there is no reference $\Delta^{2}$ value to take as a basis to compare the algorithms. Again, POUNDERS found the lowest $\Delta^{2}$ and it came out to be the most effective algorithm to parameterize the DFT dataset, although the advantage over the other two algorithms is less dramatic. However, for the same $500$ initial guesses, POUNDERS required the largest $N_{eval}=1024000$, compared to $572401$ and $107174$ for Simplex and L-M, respectively.

Figure \ref{fig:lm-vs-simplex}(c) shows the change in the minimized $\Delta^{2}$ with respect to number of initial guesses for the DFT dataset, indicating that for any number of initial guess used, POUNDERS resulted in the lowest $\Delta^{2}$, whereas Simplex and L-M have similar $\Delta^{2}$. For POUNDERS, $\Delta^{2}$ is not improved significantly when more than $250$ initial guesses are used. With L-M or Simplex, there is a slight decrease in the $\Delta^{2}$ with the number of initial guesses beyond $250$. Accordingly, although POUNDERS required larger $N_{eval}$ to reach the lowest $\Delta^{2}$, the same $\Delta^{2}$, could be achieved with the use of fewer initial guesses compared to Simplex and L-M.

The statistical count analysis of $\Delta^{2}$ values from Simplex, L-M, POUNDERS, and random search, with respect to number of initial guesses for DFT data is given in Figure \ref{fig:lm-vs-simplex}(d). Figure \ref{fig:lm-vs-simplex}(d) shows that all algorithms can reduce the value of $\Delta^{2}$ by $2-10$ orders of magnitude compared to random search of parameters. For DFT data, POUNDERS requires a larger number of initial guesses to reach $\Delta^{2}<1000$ compared to Simplex and L-M, though the difference was small.

\begin{figure}
\begin{centering}
\includegraphics[width=13cm]{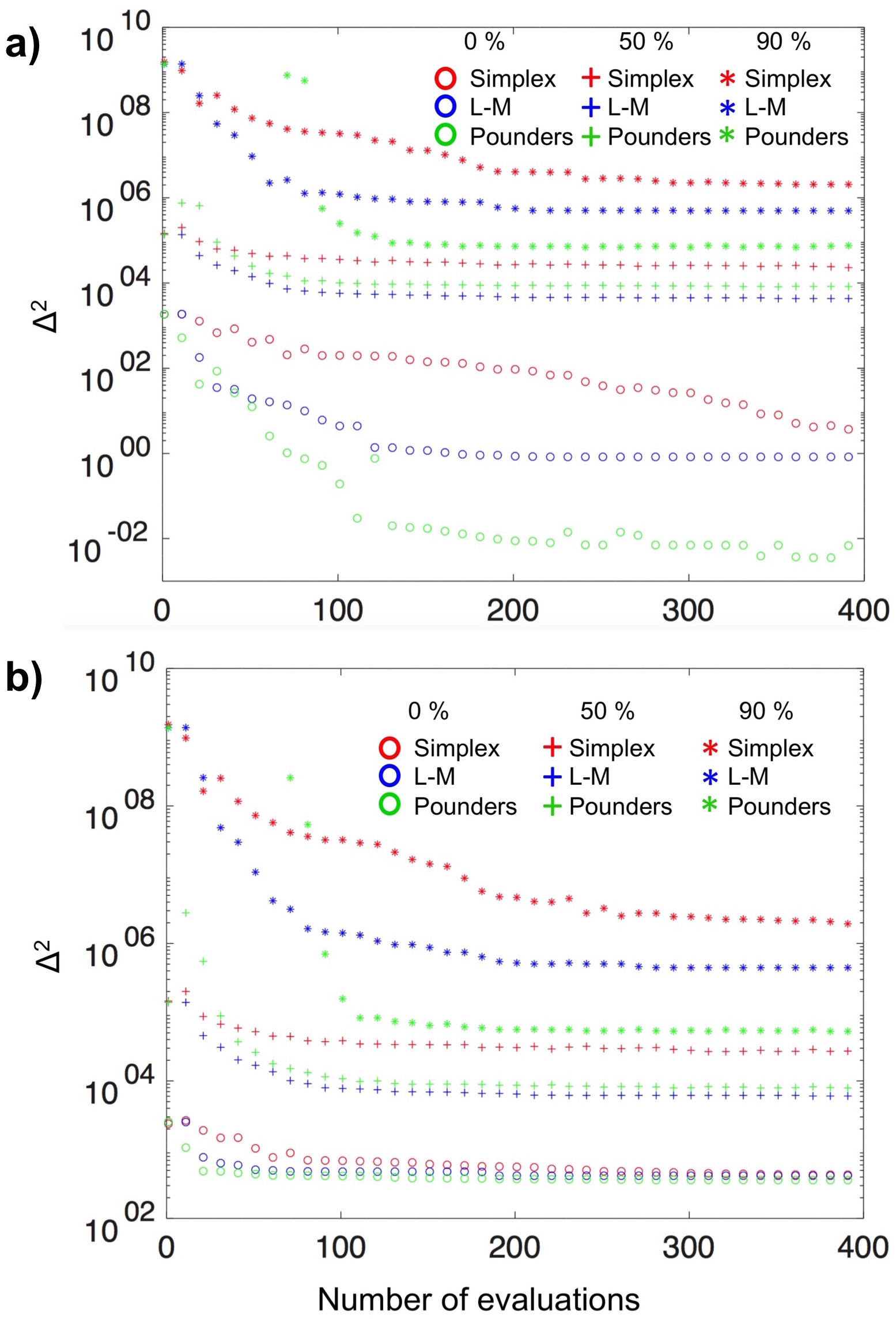}
\par\end{centering}
\protect\caption{Percentile analysis of Simplex, Levenberg-Marquardt (L-M), and POUNDERS algorithms for a) Test data and b) DFT data calculated at $0\%$, $50\%$, and $90\%$. \label{fig:percentile}}
\end{figure}

A comparison of the performance of the three algorithms in terms of how quickly they achieve the lowest solution at different starting points is carried out using percentile analysis of the $\Delta^{2}$ values. For all 500 starting guesses, we sorted the $\Delta^{2}$ calculated by Simplex, L-M, and POUNDERS at each evaluation step. Subsequently, we selected the lowest $\Delta^{2}$ ($0 \%$), and also $50 \%$-tile and $90 \%$-tile of the $\Delta^{2}$ values. The results for the Test and DFT data are plotted in Figure \ref{fig:percentile}(a) and (b), respectively. Figure \ref{fig:percentile}(a) shows that L-M reaches the lowest $0\%$ of $\Delta^{2}$ value in $N_{eval}=130$, while POUNDERS requires $N_{eval}=200$. For $0 \%$, Simplex takes the largest number of $N_{eval}=380$ to converge. For percentile values of $50 \%$, and $90 \%$, L-M required the fewest $N_{eval}$ to reach a stable $\Delta^{2}$. POUNDERS resulted in the largest drop in the $\Delta^{2}$ value both for $0 \%$, and $90 \%$, so that even if the starting guess is not close to the minimum, POUNDERS is more efficient in reducing the $\Delta^{2}$ compared to L-M and Simplex. Similar results are also observed for the DFT data in Figure \ref{fig:percentile}(b), with the exception that differences between the three methods is much smaller for the best (0\%) solution compared to the Test data. POUNDERS resulted in the largest drop in $\Delta^{2}$ for the $0 \%$, and $90 \%$, while for $0 \%$ POUNDERS uses the smallest $N_{eval}=30$ to reach a low stable $\Delta^{2}$ value. 

\subsection{Genetic algorithm (GA) results}
\subsubsection{Test data}
Again, we first determine the effectiveness of GA in finding low-$\Delta^{2}$ solutions. The evolution of $\Delta^{2}$ with the number of evaluations $(N_{eval})$ using different population sizes in SGA is shown in Figure \ref{fig:fig-sga}(a) for the Test data. Figure \ref{fig:fig-sga}(a) shows that using population size of $100$ and $400$ resulted in similar $\Delta^{2}$ value ($\approx 17$) after $50000$ evaluations. If Simplex is called at each generation using a population size of 100, $\Delta^{2}$ can be reduced to 15.6. Interestingly, population size of $1000$ could not reach to a low $\Delta^{2}$ in the same $N_{eval}$. 
For the Test data, using a population size of $400$, $\Delta^{2}$ is reduced from $17.4$ at $50000$ evaluations to $5.7$ if the simulations are continued until $200000$ evaluations. We speculate that given sufficiently large number of evaluations, the population size of $1000$ might converge to a lower $\Delta^{2}$. For Test data, using similar numbers of evaluations, SGA obtains similar $\Delta^{2}$ values to Simplex and L-M, but is less effective than POUNDERS.  

\subsubsection{DFT data}
\begin{figure}
\begin{centering}
\includegraphics[width=9cm]{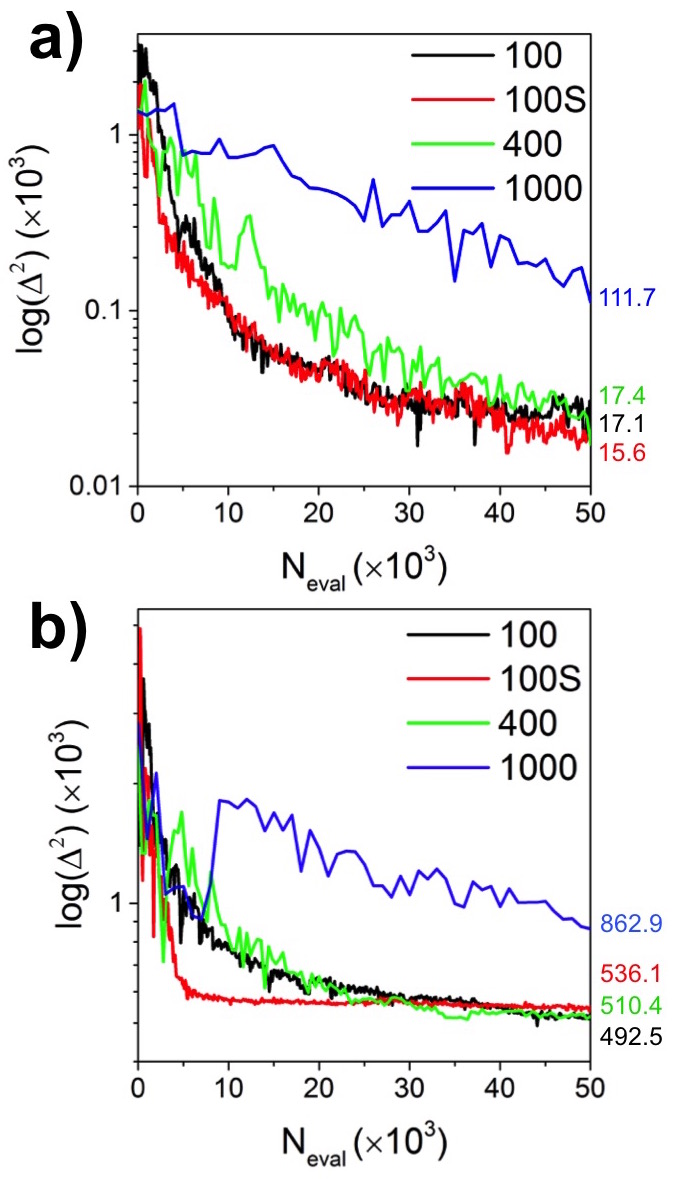}
\par\end{centering}
\protect\caption{Change in the value of $\Delta^{2}$ with number of evaluations using simple genetic algorithm (SGA) for a) Test data, and b) DFT
data. Each legend in the graphs corresponding to population size, while 100S corresponds to the case where population size of $100$ is used and Simplex is called at each generation. \label{fig:fig-sga}}
\end{figure}

In the case of DFT data, Figure \ref{fig:fig-sga}(b) shows similarly that population sizes of $100$ and $400$ result in similar values of $\Delta^{2}$ at the end of $N_{eval}=50000$, but using a population size of $1000$ results in a higher $\Delta^{2}$ for the same number of evaluations. On the other hand, when Simplex algorithm is also called at each generation for the population size of $100$, the convergence in $\Delta^{2}$ achieved in $N_{eval}<10000$. 
The lowest $\Delta^{2}=492.5$ is obtained using a population size of $100$ at the end of $50000$ evaluations, further reducing to $479.0$
if $200000$ evaluations are carried out. For DFT data, the lowest $\Delta^{2}$ values obtained from SGA are comparable to the $\Delta^{2}$ values
obtained using multi-start Simplex, but SGA reaches convergence in fewer evaluations. We also find that using a larger population requires more generations, or larger $N_{eval}$, to achieve convergence.

\subsection{Multi-objective genetic algorithm (MOGA)}
\subsubsection{Test data}
\begin{figure}
\begin{centering}
\includegraphics[width=16cm]{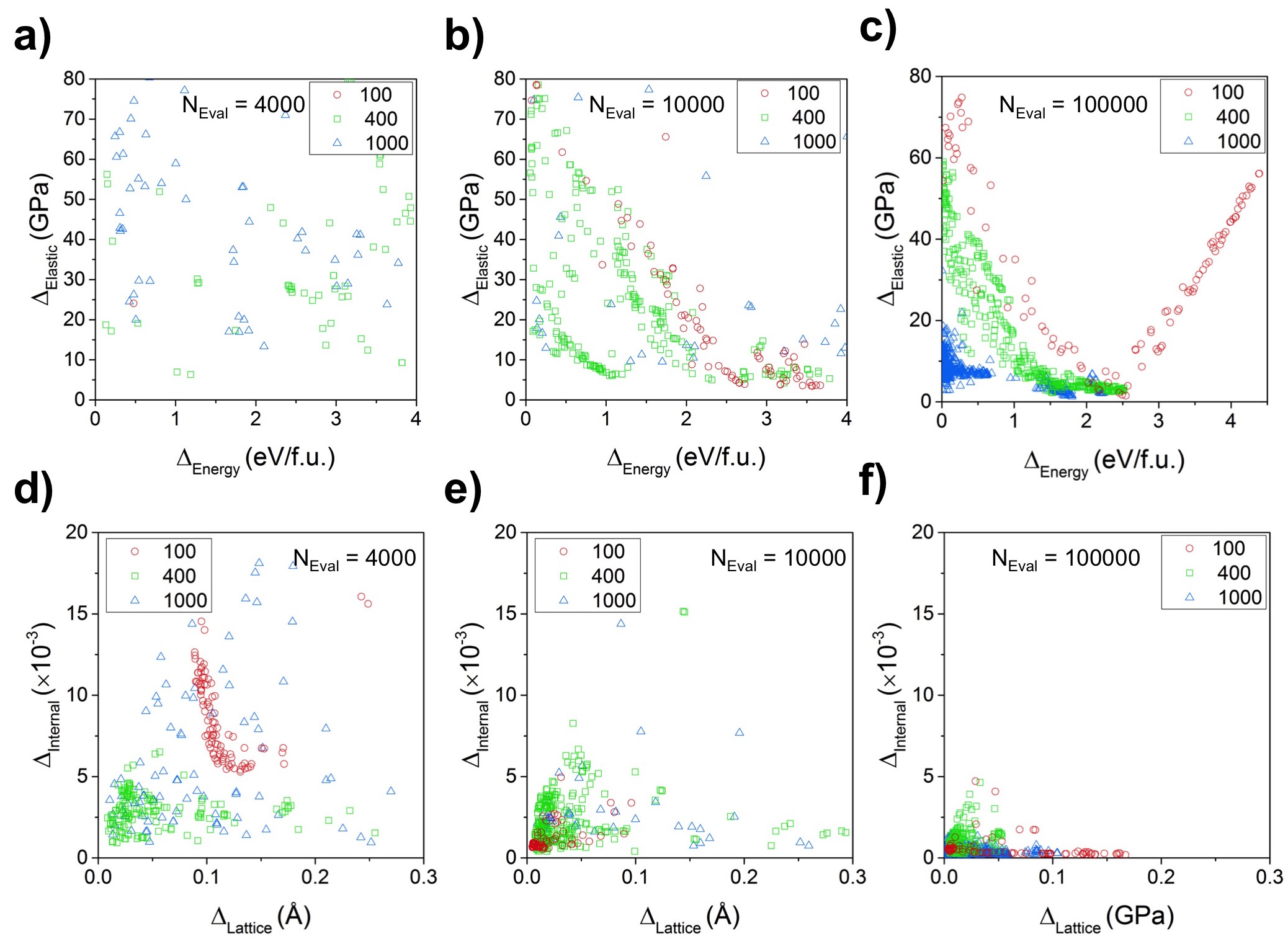}
\par\end{centering}
\protect\caption{The evolution of the RMS errors in energies and elastic constants obtained using multi-objective genetic algorithm (MOGA) on Test data. The snapshots are taken at $N_{eval}=$ a) $4000$, b) $10000$, and c) $100000$. The evolution of the RMS errors in lattice constants and internal coordinates are also shown for $N_{eval}=$ d) $4000$, e) $10000$, and f) $100000$. Calculations are carried out using population size of $100$, $400$ and $1000$ as indicated in the legend.  \label{fig:fig-moga-test-eval}}
\end{figure}

Without a single, weighted objective, evaluating the effectiveness of MOGA requires the visualization of data in various dimensions. The evolution of the root mean square (RMS) errors in energies and elastic constants with respect to number of evaluations is shown in Figure \ref{fig:fig-moga-test-eval}(a) for $N_{eval}=4000$, (b) for $N_{eval}=10000$, and (c) for $N_{eval}=100000$, using different population sizes. 
At $N_{eval}=4,000$, no Pareto front formation is observed for all the population sizes considered, while the lowest RMS errors in both energies and elastic constants are reached when the population size is $400$ (Figure \ref{fig:fig-moga-test-eval}(a)). 
Pareto front formation is observed at $N_{eval}=10000$, for both population sizes of $100$ and $400$ (Figure \ref{fig:fig-moga-test-eval}(b)).  
Beyond $N_{eval}=100000$, the Pareto front movement towards origin is stopped and the genetic algorithm is converged.  At this point, Figure \ref{fig:fig-moga-test-eval}(c) shows that the best solutions (closest to the origin) are obtained for a population of $1000$. A population size of $400$ also results in considerably small errors, but a population size of $100$ is insufficient to reach the solution. 
The change in the RMS errors in lattice constants and internal coordinates with respect to $N_{eval}$ is given in Figures \ref{fig:fig-moga-test-eval}(d)-(f)
for $N_{eval}$. At $N_{eval}=4,000$, for all the populations, there is a large spread in RMS errors in lattice constants and internal coordinates, although a Pareto front appears to form for a population size of $100$ (Figure \ref{fig:fig-moga-test-eval}(d)). When the MOGA proceeded to $N_{eval}=10000$, solutions are found closer to the origin (Figure \ref{fig:fig-moga-test-eval}(e)). Finally, at $N_{eval}=100000$, most of the solutions are collected at the zero point with very low errors in lattice constants and internal coordinates (Figure \ref{fig:fig-moga-test-eval}(f)).
When the optimization process was performed using the Test data, for which there exists an exact solution, we find that MOGA successfully arrives at solutions very close to the exact solution, when a sufficiently large population size is used.

\begin{figure}
\begin{centering}
\includegraphics[width=12cm]{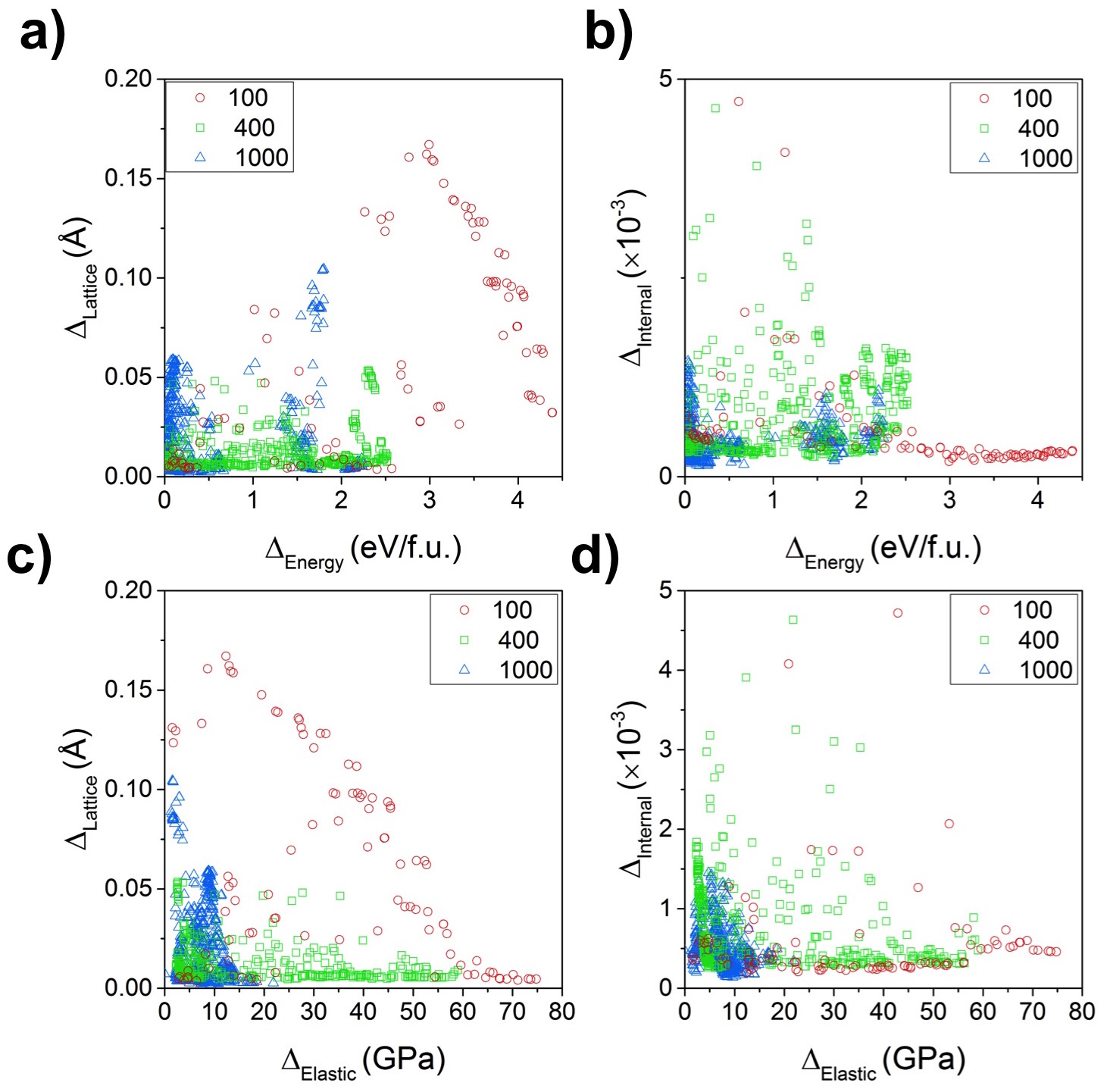}
\par\end{centering}
\protect\caption{The RMS errors for the final ($N_{eval}=100000$) solutions, obtained from MOGA, in a) lattice constants vs. energy, b) internal coordinates vs. energy, c) lattice constants vs. elastic constants, and d) internal coordinates vs. elastic constants, for population sizes of $100$, $400$, and $1000$.
\label{fig:fig-moga-test-objs}}
\end{figure}

The RMS errors for all of the quantities at $N_{eval}=100000$ are plotted in Figure \ref{fig:fig-moga-test-objs}. Figure \ref{fig:fig-moga-test-objs}(a)-(d)
shows that errors for all objectives, i.e. energies, elastic constants, lattice parameters, and internal coordinates, are converged to values very close, or equal, to zero when a sufficiently large population size is used. These results show that MOGA is an effective method for fitting force field parameters for this Test data.

\subsubsection{DFT data}
\begin{figure}
\begin{centering}
\includegraphics[width=15cm]{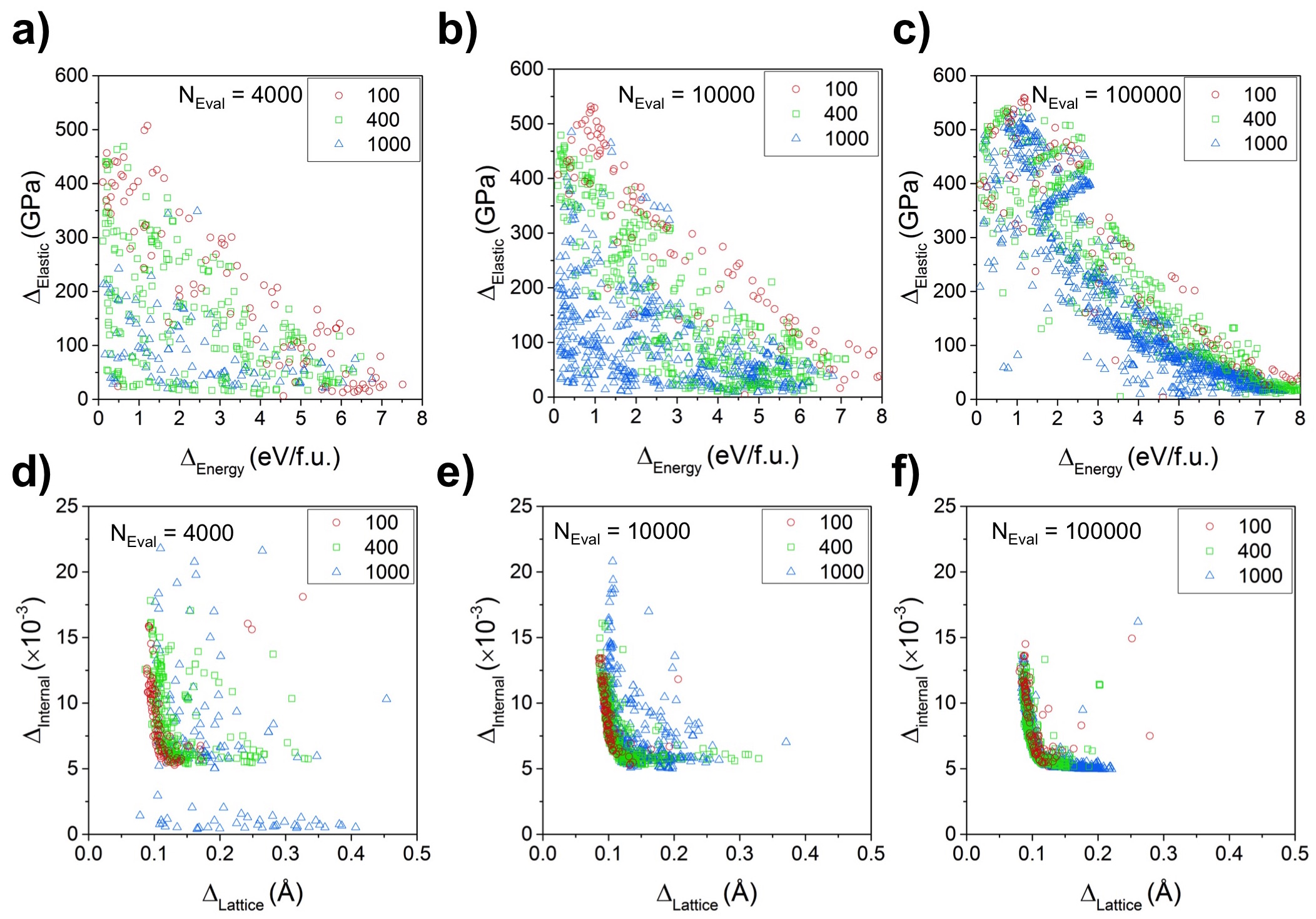}
\par\end{centering}
\protect\caption{The evolution of the RMS errors in energies and elastic constants obtained using multi-objective genetic algorithm (MOGA) with DFT data. The snapshots are taken at $N_{eval}=$ a) $4000$, b) $10000$, and c) $100000$. The evolution of the RMS errors in lattice constants and internal coordinates at $N_{eval}=$ d) $4000$, e) $10000$, and f) $100000$. Calculations are carried out using population sizes of $100$, $400$ and $1000$ as indicated in the legend. \label{fig:fig-moga-dft-eval}}
\end{figure}

The evolution of the RMS errors in energies and elastic constants with respect to number of evaluations are shown in Figures \ref{fig:fig-moga-dft-eval}(a)-(c). At $N_{eval}=4000$, using population sizes of $400$ or $1000$ results in solutions close to the origin, indicating that these two objectives allow common solutions with low errors (Figure \ref{fig:fig-moga-dft-eval}(a), but again a population of $100$ appears insufficient. 
At $N_{eval}=10000$ in Figure \ref{fig:fig-moga-dft-eval}(b), some of the solutions close to the origin disappeared for population sizes of $400$ and $1000$. On the other hand, for a population size of $100$, a Pareto front starts to develop.
Beyond $N_{eval}=100000$, the Pareto front movement towards origin is stopped and the genetic algorithm is converged. Figure \ref{fig:fig-moga-dft-eval}(c)
shows that the least errors in both energies and elastic constants are obtained with population sizes of $400$ and $1000$, as is the case for the Test data.
A population size of $100$ was not sufficient to achieve low errors. 
The RMS errors in lattice constants and internal coordinates are different, such that a Pareto front is visible starting from $N_{eval}=4000$ even for a population of 100, and  
by $N_{eval}=100000$ (Figure \ref{fig:fig-moga-dft-eval}(f)), most of the solutions are at the Pareto front regardless of the population size. Unlike the Test data case in Figure \ref{fig:fig-moga-test-eval}, solutions that have better energies and elastic constants are eliminated in order to obtain a Pareto front for all four of the objectives simultaneously for DFT data as seen in Figure \ref{fig:fig-moga-dft-eval}(a) and (c).

\begin{figure}
\begin{centering}
\includegraphics[width=12cm]{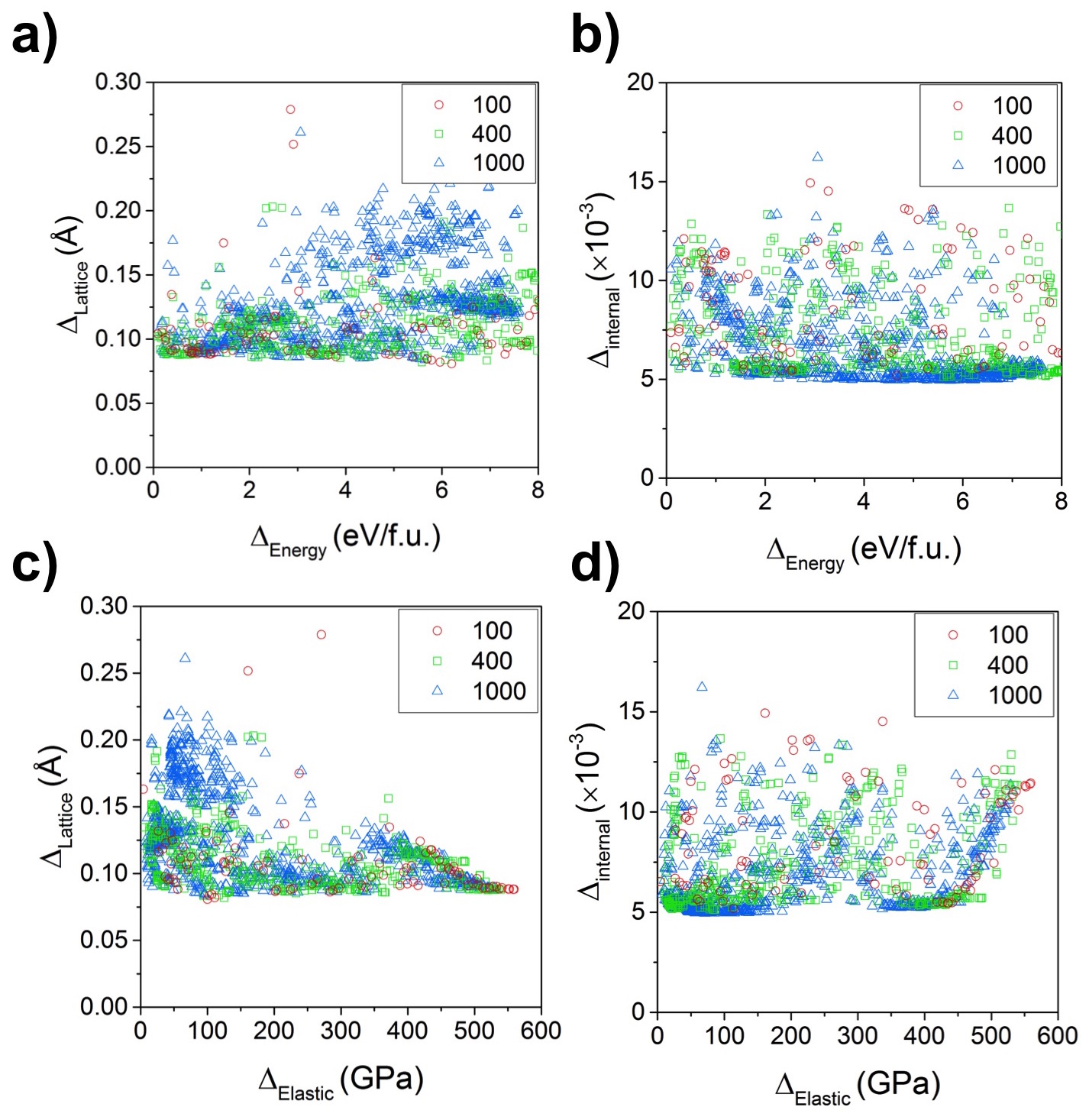}
\par\end{centering}
\protect\caption{The final ($N_{eval}=100000$) RMS errors in a) lattice constants vs. energy, b) internal coordinates vs. energy, c) lattice constants vs. elastic constants, and d) internal coordinates vs. elastic constants. All snapshots are obtained at the end of $100000$ evaluations using MOGA for the DFT data with different population sizes of $100$, $400$, and $1000$.
\label{fig:fig-moga-dft-objs}}
\end{figure}

\begin{figure}
\begin{centering}
\includegraphics[width=8cm]{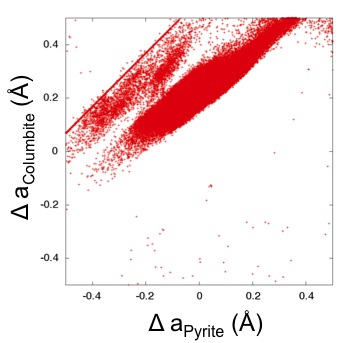}
\par\end{centering}
\protect\caption{The errors in the lattice parameter ($a$) of pyrite and columbite IrO$_2$ with respect to DFT data for all of the evaluations obtained using
population size of $1000$ in MOGA. \label{fig:fig-a-pyrite-vs-columbite}}
\end{figure}

The evolution of solutions are plotted in binary combinations in  Figures \ref{fig:fig-moga-dft-objs}(a)-(d). We see that no solution is available below a certain limit for the RMS error in lattice constants and internal coordinates -- the RMS error in lattice constants cannot be reduced below $0.08$ and the RMS
error in internal coordinates can not be reduced below $0.005$. This behavior likely indicates the presence of conflicting constraints in the training set. 
To understand the asymptotic behavior in the errors of lattice constants and internal coordinates, we analyze all the errors in individual lattice constant values in different crystal structures. We find that FF parameters that accurately predict the pyrite lattice parameter overestimate the columbite lattice parameter, and vice versa, so that the total errors in lattice constants for the pyrite and the columbite phases can not be reduced below a certain
value as shown in Figure \ref{fig:fig-a-pyrite-vs-columbite}. 
Similar results are obtained for the internal coordinates of these structures. On the other hand, the errors in lattice constants of other structures can be simultaneously very low such that the many solutions are in the proximity of the origin. From these results, one infers that the pairwise Morse FF
is not sophisticated enough to predict the structures of pyrite and the columbite phases concurrently. Hence, we propose that multi-objective optimization methods are suitable for testing the transferability of a FF.

\section{Discussion}
\subsection{Genetic algorithm vs multi-start local schemes}
The main difference between the local search and the global optimization methods is the dependence of local search algorithms on the initial guesses. 
Using the Test data, we find that Simplex, L-M, and POUNDERS reach a lower $\Delta^{2}$ value of $2.97$, $0.83$ and $0.003$, respectively, compared to
single-objective GA with $\Delta^{2}=17.4$ if a population size of $400$ was used.
Figure \ref{fig:lm-vs-simplex}(b) shows that only $1.8\%$ ($7$) of the initial guesses lead to $\Delta^{2}<17.4$ for L-M and $<1\%$ ($4$) of the initial guesses lead to $\Delta^{2}<17.4$ for the Simplex method, whereas $7.6\%$ ($38$) of the initial guesses lead to $\Delta^{2}<17.4$ if POUNDERS  is used. 
The difference in the efficiency is similar for the DFT data, in which Simplex, L-M, and POUNDERS find $\Delta^{2}$ values of $432$, $419$, and $353$, respectively, which are lower than $\Delta^{2}=493$ reached by single-objective GA. 
However, with Simplex and L-M, only $<1\%$ ($3,2$) of the initial guesses give a lower $\Delta^{2}$ compared to GA as shown in Figure \ref{fig:lm-vs-simplex}(d). On the other hand, when POUNDERS is used, $7.8\%$ ($39$) of initial guesses reached better solution than GA. 
Consequently, in order to find the global minimum of the parameter space, one should sample with a large number of initial guesses if a local search method
is used, especially Simplex and L-M. Alternatively, GA appears to be more efficient and gives reasonably accurate solution. 
In our calculations, if we carry out longer GA runs with $500$ generations using a population size of $400$, which makes $N_{eval}=200000$, we could reduce the $\Delta^{2}$ to $5.9$ and $479$ in test and DFT data, respectively. These values are very close to the lowest $\Delta^{2}$ values obtained with local search methods and still required fewer function evaluations than POUNDERS and Simplex computed with $500$ initial guesses. 
The best of the both worlds is combining both local and global optimization methods. Figure \ref{fig:fig-sga}(b) shows that the $\Delta^{2}$ is converged to the lowest point at faster (i.e. smaller $N_{eval}$) for DFT data when Simplex and GA are coupled.

\subsection{Single vs multi-objective genetic algorithm}
The most prominent advantage of using a multi-objective algorithm, vs. a single-objective one, is to remove the ambiguities brought by the weighting of fitted quantities. In MOGA, all the four objectives, namely energies, elastic constants, lattice constants and internal coordinates are optimized simultaneously to obtain the Pareto front.
To compare the efficiency and accuracy of MOGA with respect to SGA, we weight all the solutions of MOGA at each generation, and recorded the lowest $\Delta^{2}$ value at each generation. Accordingly, we have the same weighted $\Delta^{2}$ value comparable with SGA. For the Test data, MOGA reaches $\Delta^{2}$ of $7.09$ at the end of $N_{eval}=100000$, which is even smaller than the $\Delta^{2}=12.7$ obtained at the same $N_{eval}$ for SGA.
Consequently, we find that MOGA can also reach the same solution as single objective GA, but with a great advantage of the absence of weights. One can
easily weigh the objectives after the optimization according to the problem and the desired properties.

In single-objective GA, although we attempted to weight the different category of errors approximately equally, unexpected constraints in the data set dictates otherwise.  
We find that at the lowest $\Delta^{2}=479.0$, the contributions of energies, elastic constants, lattice constants, and internal coordinates towards $\Delta^2$ are $6.1$, $44.2$, $278.9$, and $149.7$, respectively. Accordingly, $\Delta^{2}$ is mostly dominated by the errors in lattice parameters and internal coordinates, likely due to the conflict between the pyrite and columbite phases described above, whereas errors in energy contributes
only $\approx 1\%$ of the total $\Delta^{2}$. This again illustrates potential pitfalls when parameterizing FFs to reproduce different quantities using single-objective optimization. 

\begin{figure}
\begin{centering}
\includegraphics[width=12cm]{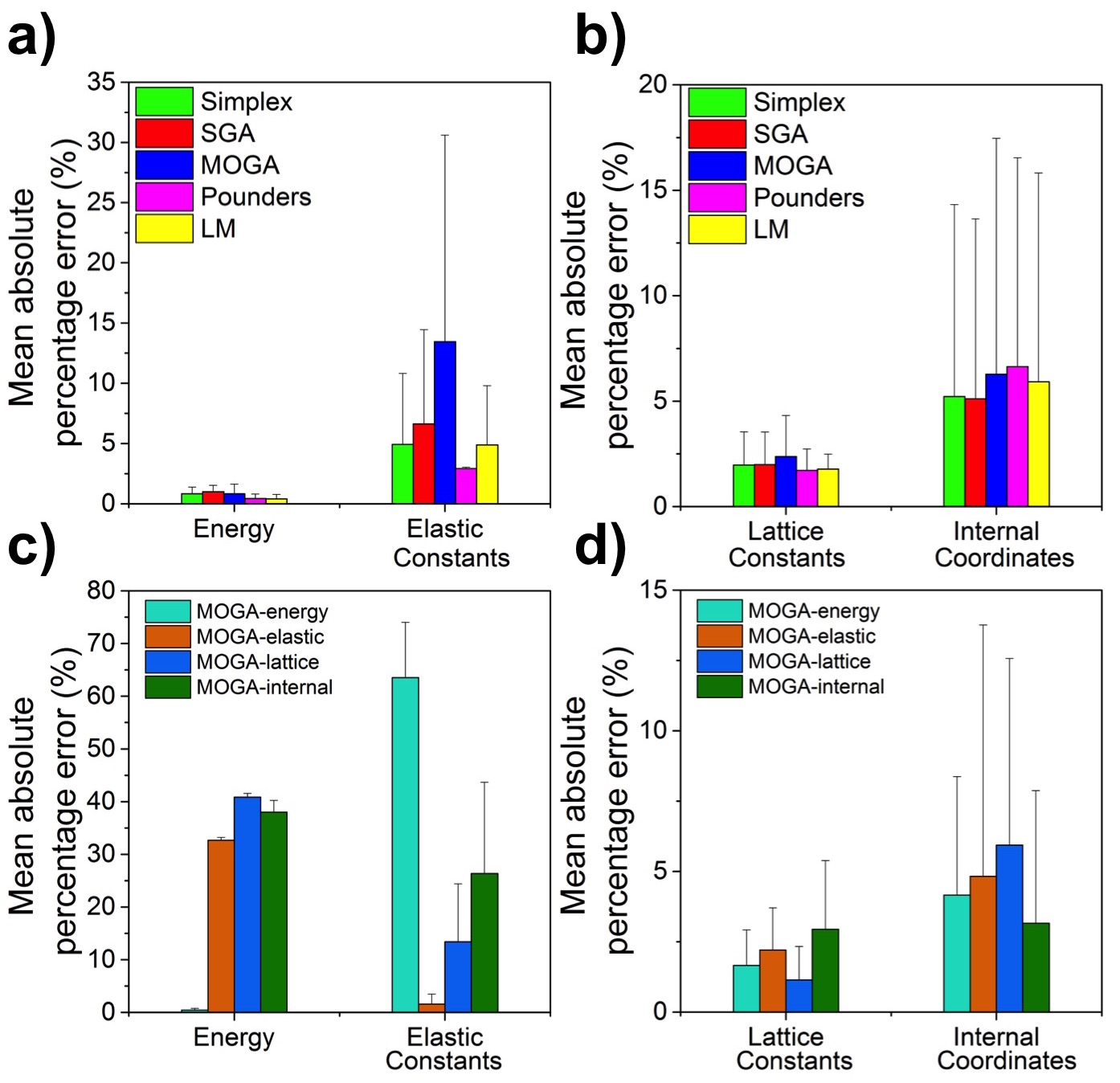}
\par\end{centering}
\protect\caption{Mean absolute percentage errors in Simplex, L-M, POUNDERS, SGA and MOGA. \label{fig:fig-map-moga-vs-sga}}
\end{figure}

Figure \ref{fig:fig-map-moga-vs-sga} shows the mean absolute percentage errors in different observables compared to the DFT data using Simplex, L-M, POUNDERS, SGA (population $=400$), and MOGA (population $=1000$).  
Since MOGA gives a set of solutions, we select a relatively optimum solution by first capturing all the solutions that have RMS error in energy $<1.0$ eV.
In the remaining set, we select the solutions that have RMS error in elastic constants $<35$ GPa. Then from the remaining solutions, we select a solution
that has low errors in lattice constants and internal coordinates.
In addition to this solution, in Figure \ref{fig:fig-map-moga-vs-sga} we include the solutions that have the lowest error in energy (MOGA-energy), lowest error in elastic constants (MOGA-elastic), lowest error in lattice constants (MOGA-lattice), and lowest error in internal coordinates (MOGA-internal). 
Figure \ref{fig:fig-map-moga-vs-sga}(a) and (b) shows that POUNDERS has lower or similar errors in all four observables than SGA. 
However, comparison of POUNDERS and SGA with MOGA is not straightforward. For instance, MOGA-energy has the lowest error in energy, lattice constants and internal coordinates, but this solution gives the largest error in elastic constants (Figure \ref{fig:fig-map-moga-vs-sga}(c) and (d)). 
Similarly, MOGA-elastic gives the lowest error in elastic constants but it produces large errors in energy values (Figure \ref{fig:fig-map-moga-vs-sga}(c)). If one wants to get the lowest error in lattice constants or internal coordinates, MOGA-lattice and MOGA-internal are the best solutions (Figure \ref{fig:fig-map-moga-vs-sga}(d)). This comparison simultaneously illustrates the possibility of tailored FFs for different properties available from MOGA, and the inherent difficulty in balancing the accurate prediction of different properties for a generally-applicable FF.

\section{Conclusions}
A comparison of different local and global optimization algorithms for parameterizing force fields (FFs) is carried out using both a test data set generated from a known FF, and a DFT data set. The goal is to reproduce target values of energies, elastic constants, lattice constants, and internal co-ordinates for various  Ir$O_{2}$ crystal polymorphs. For single-objective optimization, we use a weighted combined objective encompassing all target value types, whereas for multi-objective optimization, these different categories are treated separately in the optimization procedure. We find that:
\begin{itemize}
    \item One important criterion in evaluating the optimization strategies is the effectiveness, i.e. the ability of reaching FFs with low predictive errors compared to the target. For the Test data set, POUNDERS is by far the most effective in reaching the known lowest-error solution, compared to Simplex, Levenberg-Marquardt (L-M), and single- and multi-objective genetic algorithm (GA). 
    For the DFT data set, though POUNDERS also obtained the lowest values for the objective function, the effectiveness of the different optimization approaches is much more similar, with single-objective GA being more efficient in terms of producing similar comparable solutions with the fewest number of evaluations. The fact that the effectiveness of an optimization approach in finding the optimal solution is drastically different between the Test and DFT data set is an important lesson, highlighting the need to test optimization algorithms using realistic rather than idealized data.
    \item Another consideration is the efficiency of the optimization strategies, i.e. the number of function evaluations needed to obtain a good solution. We find that POUNDERS requires $\sim 10$ times more function evaluations than L-M and $\sim 2$ times more function evaluations than Simplex for each initial guess. 
    Compared to local optimization, genetic algorithm (GA) is found to be a more efficient method to obtain a reasonable solution, requiring a smaller number of function evaluations than multi-start local optimization. Furthermore, POUNDERS, Simplex and L-M algorithms are strongly dependent on the choice and the number of initial guess points. The success rate of reaching a similar solution as compared to GA is less than 1\% for Simplex and L-M, while $8\%$ of initial guesses could reach lower solution than GA if POUNDERS is used. Even faster convergence can be achieved with GA if it is coupled with the Simplex method. Note that the chosen FF only has 9 parameters, whereas typical FFs would have significantly more, in which case one could expect mutli-start local optimization to require more initial guesses. 
    \item Weighting the different types of predicted properties can pose significant difficulties. In the current study, using pre-determined weights representative of the uncertainties in the DFT predictions in energies, elastic constants, lattice parameter, and internal co-ordinates, we nonetheless find that errors in the latter two categories dominate due to a conflict between two phases. 
    The use of multi-objective GA (MOGA) can eliminate the ambiguities and difficulties due to weighting scheme. MOGA reproduces the same solution as  single-objective GA, when parameterization is carried out using the Test data set. MOGA gives as output a number of possible parameter sets, and we show that a reasonable "compromise" solution is comparable to those obtained from single-objective optimization. However, MOGA provides in addition parameter sets that excel in specific properties while still giving reasonable accuracy in others, making it possible to tailor FFs for different properties, if at the expense of transferability.
\end{itemize}


\section*{Acknowledgment}
This material is based upon work supported by Laboratory Directed Research and Development (LDRD) funding from Argonne National Laboratory, provided by the Director, Office of Science, of the U.S. Department of Energy under Contract No. DE-AC02-06CH11357 Use of the Center for Nanoscale Materials, an Office of Science user facility, was supported by the U.S. Department of Energy, Office of Science, Office of Basic Energy Sciences, under Contract No. DE-AC02-06CH11357. We gratefully acknowledge the computing resources provided on Blues, a high-performance computing cluster operated by the Laboratory Computing Resource Center at Argonne National Laboratory. 

\bibliographystyle{apsrev}
\bibliography{refs}

\begin{thebibliography}{81}
\expandafter\ifx\csname natexlab\endcsname\relax\def\natexlab#1{#1}\fi
\expandafter\ifx\csname bibnamefont\endcsname\relax
  \def\bibnamefont#1{#1}\fi
\expandafter\ifx\csname bibfnamefont\endcsname\relax
  \def\bibfnamefont#1{#1}\fi
\expandafter\ifx\csname citenamefont\endcsname\relax
  \def\citenamefont#1{#1}\fi
\expandafter\ifx\csname url\endcsname\relax
  \def\url#1{\texttt{#1}}\fi
\expandafter\ifx\csname urlprefix\endcsname\relax\def\urlprefix{URL }\fi
\providecommand{\bibinfo}[2]{#2}
\providecommand{\eprint}[2][]{\url{#2}}

\bibitem[{\citenamefont{Abraham et~al.}(2002)\citenamefont{Abraham, Walkup,
  Gao, Duchaineau, De~La~Rubia, and Seager}}]{Abraham2002}
\bibinfo{author}{\bibfnamefont{F.~F.} \bibnamefont{Abraham}},
  \bibinfo{author}{\bibfnamefont{R.}~\bibnamefont{Walkup}},
  \bibinfo{author}{\bibfnamefont{H.}~\bibnamefont{Gao}},
  \bibinfo{author}{\bibfnamefont{M.}~\bibnamefont{Duchaineau}},
  \bibinfo{author}{\bibfnamefont{T.~D.} \bibnamefont{De~La~Rubia}},
  \bibnamefont{and} \bibinfo{author}{\bibfnamefont{M.}~\bibnamefont{Seager}},
  \bibinfo{journal}{Proceedings of the National Academy of Sciences}
  \textbf{\bibinfo{volume}{99}}, \bibinfo{pages}{5777} (\bibinfo{year}{2002}).

\bibitem[{\citenamefont{Kadau et~al.}(2002)\citenamefont{Kadau, Germann,
  Lomdahl, and Holian}}]{Kadau2002}
\bibinfo{author}{\bibfnamefont{K.}~\bibnamefont{Kadau}},
  \bibinfo{author}{\bibfnamefont{T.~C.} \bibnamefont{Germann}},
  \bibinfo{author}{\bibfnamefont{P.~S.} \bibnamefont{Lomdahl}},
  \bibnamefont{and} \bibinfo{author}{\bibfnamefont{B.~L.}
  \bibnamefont{Holian}}, \bibinfo{journal}{Science}
  \textbf{\bibinfo{volume}{296}}, \bibinfo{pages}{1681} (\bibinfo{year}{2002}).

\bibitem[{\citenamefont{Vashishta et~al.}(2007)\citenamefont{Vashishta, Kalia,
  Nakano, Homan, and McNesby}}]{Vashishta2008}
\bibinfo{author}{\bibfnamefont{P.}~\bibnamefont{Vashishta}},
  \bibinfo{author}{\bibfnamefont{R.~K.} \bibnamefont{Kalia}},
  \bibinfo{author}{\bibfnamefont{A.}~\bibnamefont{Nakano}},
  \bibinfo{author}{\bibfnamefont{B.~E.} \bibnamefont{Homan}}, \bibnamefont{and}
  \bibinfo{author}{\bibfnamefont{K.~L.} \bibnamefont{McNesby}},
  \bibinfo{journal}{J. Propulsion and Power} \textbf{\bibinfo{volume}{23}},
  \bibinfo{pages}{688} (\bibinfo{year}{2007}).

\bibitem[{\citenamefont{Shekhar et~al.}(2013)\citenamefont{Shekhar, Nomura,
  Kalia, Nakano, and Vashishta}}]{Shekhar2013}
\bibinfo{author}{\bibfnamefont{A.}~\bibnamefont{Shekhar}},
  \bibinfo{author}{\bibfnamefont{K.-i.} \bibnamefont{Nomura}},
  \bibinfo{author}{\bibfnamefont{R.~K.} \bibnamefont{Kalia}},
  \bibinfo{author}{\bibfnamefont{A.}~\bibnamefont{Nakano}}, \bibnamefont{and}
  \bibinfo{author}{\bibfnamefont{P.}~\bibnamefont{Vashishta}},
  \bibinfo{journal}{Phys. Rev. Lett.} \textbf{\bibinfo{volume}{111}},
  \bibinfo{pages}{184503} (\bibinfo{year}{2013}).

\bibitem[{\citenamefont{Berman et~al.}(2015)\citenamefont{Berman, Deshmukh,
  Sankaranarayanan, Erdemir, and Sumant}}]{Berman2015}
\bibinfo{author}{\bibfnamefont{D.}~\bibnamefont{Berman}},
  \bibinfo{author}{\bibfnamefont{S.~A.} \bibnamefont{Deshmukh}},
  \bibinfo{author}{\bibfnamefont{S.~K. R.~S.} \bibnamefont{Sankaranarayanan}},
  \bibinfo{author}{\bibfnamefont{A.}~\bibnamefont{Erdemir}}, \bibnamefont{and}
  \bibinfo{author}{\bibfnamefont{A.~V.} \bibnamefont{Sumant}},
  \bibinfo{journal}{Science} \textbf{\bibinfo{volume}{348}},
  \bibinfo{pages}{1118} (\bibinfo{year}{2015}).

\bibitem[{\citenamefont{Allen and Tildesley}(1989)}]{Allen1989computer}
\bibinfo{author}{\bibfnamefont{M.~P.} \bibnamefont{Allen}} \bibnamefont{and}
  \bibinfo{author}{\bibfnamefont{D.~J.} \bibnamefont{Tildesley}},
  \emph{\bibinfo{title}{Computer simulation of liquids}}
  (\bibinfo{publisher}{Oxford university press}, \bibinfo{year}{1989}).

\bibitem[{\citenamefont{Tersoff}(1988)}]{Tersoff1988}
\bibinfo{author}{\bibfnamefont{J.}~\bibnamefont{Tersoff}},
  \bibinfo{journal}{Phys. Rev. B} \textbf{\bibinfo{volume}{37}},
  \bibinfo{pages}{6991} (\bibinfo{year}{1988}).

\bibitem[{\citenamefont{Daw et~al.}(1993)\citenamefont{Daw, Foiles, and
  Baskes}}]{Daw1993eam}
\bibinfo{author}{\bibfnamefont{M.~S.} \bibnamefont{Daw}},
  \bibinfo{author}{\bibfnamefont{S.~M.} \bibnamefont{Foiles}},
  \bibnamefont{and} \bibinfo{author}{\bibfnamefont{M.~I.}
  \bibnamefont{Baskes}}, \bibinfo{journal}{Mater. Sci. Reports}
  \textbf{\bibinfo{volume}{9}}, \bibinfo{pages}{251} (\bibinfo{year}{1993}).

\bibitem[{\citenamefont{Bush et~al.}(1994)\citenamefont{Bush, Gale, Catlow, and
  Battle}}]{Catlow1994}
\bibinfo{author}{\bibfnamefont{T.~S.} \bibnamefont{Bush}},
  \bibinfo{author}{\bibfnamefont{J.~D.} \bibnamefont{Gale}},
  \bibinfo{author}{\bibfnamefont{C.~R.~A.} \bibnamefont{Catlow}},
  \bibnamefont{and} \bibinfo{author}{\bibfnamefont{P.~D.}
  \bibnamefont{Battle}}, \bibinfo{journal}{J. Mater. Chem.}
  \textbf{\bibinfo{volume}{4}}, \bibinfo{pages}{831} (\bibinfo{year}{1994}).

\bibitem[{\citenamefont{Finnis}(2007)}]{Finnis2007bop}
\bibinfo{author}{\bibfnamefont{M.~W.} \bibnamefont{Finnis}},
  \bibinfo{journal}{Prog. Mater. Sci.} \textbf{\bibinfo{volume}{52}},
  \bibinfo{pages}{133} (\bibinfo{year}{2007}).

\bibitem[{\citenamefont{Sinnott and Brenner}(2012)}]{Sinnott2012}
\bibinfo{author}{\bibfnamefont{S.~B.} \bibnamefont{Sinnott}} \bibnamefont{and}
  \bibinfo{author}{\bibfnamefont{D.~W.} \bibnamefont{Brenner}},
  \bibinfo{journal}{MRS Bull.} \textbf{\bibinfo{volume}{37}},
  \bibinfo{pages}{469} (\bibinfo{year}{2012}).

\bibitem[{\citenamefont{Ercolessi and Adams}(1994)}]{Ercolessi1994}
\bibinfo{author}{\bibfnamefont{F.}~\bibnamefont{Ercolessi}} \bibnamefont{and}
  \bibinfo{author}{\bibfnamefont{J.~B.} \bibnamefont{Adams}},
  \bibinfo{journal}{EPL (Europhys. Lett.)} \textbf{\bibinfo{volume}{26}},
  \bibinfo{pages}{583} (\bibinfo{year}{1994}).

\bibitem[{\citenamefont{Gale and Rohl}(2003)}]{Gale03}
\bibinfo{author}{\bibfnamefont{J.~D.} \bibnamefont{Gale}} \bibnamefont{and}
  \bibinfo{author}{\bibfnamefont{A.~L.} \bibnamefont{Rohl}},
  \bibinfo{journal}{Mol. Simul.} \textbf{\bibinfo{volume}{29}},
  \bibinfo{pages}{291} (\bibinfo{year}{2003}), ISSN \bibinfo{issn}{0892-7022}.

\bibitem[{\citenamefont{Waldher et~al.}(2010)\citenamefont{Waldher, Kuta, Chen,
  Henson, and Clark}}]{Waldher2010forcefit}
\bibinfo{author}{\bibfnamefont{B.}~\bibnamefont{Waldher}},
  \bibinfo{author}{\bibfnamefont{J.}~\bibnamefont{Kuta}},
  \bibinfo{author}{\bibfnamefont{S.}~\bibnamefont{Chen}},
  \bibinfo{author}{\bibfnamefont{N.}~\bibnamefont{Henson}}, \bibnamefont{and}
  \bibinfo{author}{\bibfnamefont{A.~E.} \bibnamefont{Clark}},
  \bibinfo{journal}{J. Comput. Chem.} \textbf{\bibinfo{volume}{31}},
  \bibinfo{pages}{2307} (\bibinfo{year}{2010}).

\bibitem[{\citenamefont{Artrith and Kolpak}(2014)}]{Artrith2014neural}
\bibinfo{author}{\bibfnamefont{N.}~\bibnamefont{Artrith}} \bibnamefont{and}
  \bibinfo{author}{\bibfnamefont{A.~M.} \bibnamefont{Kolpak}},
  \bibinfo{journal}{Nano Lett.} \textbf{\bibinfo{volume}{14}},
  \bibinfo{pages}{2670} (\bibinfo{year}{2014}).

\bibitem[{\citenamefont{Grimme}(2014)}]{Grimme2014qmdff}
\bibinfo{author}{\bibfnamefont{S.}~\bibnamefont{Grimme}}, \bibinfo{journal}{J.
  Chem. Theory Comput.} \textbf{\bibinfo{volume}{10}}, \bibinfo{pages}{4497}
  (\bibinfo{year}{2014}).

\bibitem[{\citenamefont{Jaramillo-Botero
  et~al.}(2014)\citenamefont{Jaramillo-Botero, Naserifar, and
  Goddard~III}}]{Jaramillo2014garffield}
\bibinfo{author}{\bibfnamefont{A.}~\bibnamefont{Jaramillo-Botero}},
  \bibinfo{author}{\bibfnamefont{S.}~\bibnamefont{Naserifar}},
  \bibnamefont{and} \bibinfo{author}{\bibfnamefont{W.~A.}
  \bibnamefont{Goddard~III}}, \bibinfo{journal}{J. Chem. Theory Comput.}
  \textbf{\bibinfo{volume}{10}}, \bibinfo{pages}{1426} (\bibinfo{year}{2014}).

\bibitem[{\citenamefont{Thompson et~al.}(2014)\citenamefont{Thompson, Meredig,
  Stan, and Wolverton}}]{Thompson2014}
\bibinfo{author}{\bibfnamefont{A.~E.} \bibnamefont{Thompson}},
  \bibinfo{author}{\bibfnamefont{B.}~\bibnamefont{Meredig}},
  \bibinfo{author}{\bibfnamefont{M.}~\bibnamefont{Stan}}, \bibnamefont{and}
  \bibinfo{author}{\bibfnamefont{C.}~\bibnamefont{Wolverton}},
  \bibinfo{journal}{J. Nucl. Mater.} \textbf{\bibinfo{volume}{446}},
  \bibinfo{pages}{155} (\bibinfo{year}{2014}).

\bibitem[{\citenamefont{Wang et~al.}(2014)\citenamefont{Wang, Martinez, and
  Pande}}]{Wang2014FB}
\bibinfo{author}{\bibfnamefont{L.-P.} \bibnamefont{Wang}},
  \bibinfo{author}{\bibfnamefont{T.~J.} \bibnamefont{Martinez}},
  \bibnamefont{and} \bibinfo{author}{\bibfnamefont{V.~S.} \bibnamefont{Pande}},
  \bibinfo{journal}{J. Phys. Chem. Lett.} \textbf{\bibinfo{volume}{5}},
  \bibinfo{pages}{1885} (\bibinfo{year}{2014}).

\bibitem[{\citenamefont{Jones}(1924)}]{Jones1924}
\bibinfo{author}{\bibfnamefont{J.~E.} \bibnamefont{Jones}}, in
  \emph{\bibinfo{booktitle}{Proc. Royal Soc. London A}}
  (\bibinfo{organization}{The Royal Society}, \bibinfo{year}{1924}), vol.
  \bibinfo{volume}{106}, pp. \bibinfo{pages}{463--477}.

\bibitem[{\citenamefont{Buckingham}(1938)}]{Buckingham1938}
\bibinfo{author}{\bibfnamefont{R.~A.} \bibnamefont{Buckingham}}, in
  \emph{\bibinfo{booktitle}{Proc. Royal Soc. London A}}
  (\bibinfo{organization}{The Royal Society}, \bibinfo{year}{1938}), vol.
  \bibinfo{volume}{168}, pp. \bibinfo{pages}{264--283}.

\bibitem[{\citenamefont{Finnis and Sinclair}(1984)}]{Finnis1984}
\bibinfo{author}{\bibfnamefont{M.~W.} \bibnamefont{Finnis}} \bibnamefont{and}
  \bibinfo{author}{\bibfnamefont{J.~E.} \bibnamefont{Sinclair}},
  \bibinfo{journal}{Phil. Mag. A} \textbf{\bibinfo{volume}{50}},
  \bibinfo{pages}{45} (\bibinfo{year}{1984}).

\bibitem[{\citenamefont{Stillinger and Weber}(1985)}]{Stillinger1985}
\bibinfo{author}{\bibfnamefont{F.~H.} \bibnamefont{Stillinger}}
  \bibnamefont{and} \bibinfo{author}{\bibfnamefont{T.~A.} \bibnamefont{Weber}},
  \bibinfo{journal}{Phys. Rev. B} \textbf{\bibinfo{volume}{31}},
  \bibinfo{pages}{5262} (\bibinfo{year}{1985}).

\bibitem[{\citenamefont{Baskes}(1992)}]{Baskes1992meam}
\bibinfo{author}{\bibfnamefont{M.~I.} \bibnamefont{Baskes}},
  \bibinfo{journal}{Phys. Rev. B} \textbf{\bibinfo{volume}{46}},
  \bibinfo{pages}{2727} (\bibinfo{year}{1992}).

\bibitem[{\citenamefont{Brenner}(1990)}]{Brenner1990bop}
\bibinfo{author}{\bibfnamefont{D.~W.} \bibnamefont{Brenner}},
  \bibinfo{journal}{Phys. Rev. B} \textbf{\bibinfo{volume}{42}},
  \bibinfo{pages}{9458} (\bibinfo{year}{1990}).

\bibitem[{\citenamefont{Brenner et~al.}(2002)\citenamefont{Brenner, Shenderova,
  Harrison, Stuart, Ni, and Sinnott}}]{Brenner2002rebo}
\bibinfo{author}{\bibfnamefont{D.~W.} \bibnamefont{Brenner}},
  \bibinfo{author}{\bibfnamefont{O.~A.} \bibnamefont{Shenderova}},
  \bibinfo{author}{\bibfnamefont{J.~A.} \bibnamefont{Harrison}},
  \bibinfo{author}{\bibfnamefont{S.~J.} \bibnamefont{Stuart}},
  \bibinfo{author}{\bibfnamefont{B.}~\bibnamefont{Ni}}, \bibnamefont{and}
  \bibinfo{author}{\bibfnamefont{S.~B.} \bibnamefont{Sinnott}},
  \bibinfo{journal}{J. Phys. Cond. Matter} \textbf{\bibinfo{volume}{14}},
  \bibinfo{pages}{783} (\bibinfo{year}{2002}).

\bibitem[{\citenamefont{Van~Duin et~al.}(2001)\citenamefont{Van~Duin, Dasgupta,
  Lorant, and Goddard}}]{vanDuin2001reaxff}
\bibinfo{author}{\bibfnamefont{A.~C.~T.} \bibnamefont{Van~Duin}},
  \bibinfo{author}{\bibfnamefont{S.}~\bibnamefont{Dasgupta}},
  \bibinfo{author}{\bibfnamefont{F.}~\bibnamefont{Lorant}}, \bibnamefont{and}
  \bibinfo{author}{\bibfnamefont{W.~A.} \bibnamefont{Goddard}},
  \bibinfo{journal}{J. Phys. Chem. A} \textbf{\bibinfo{volume}{105}},
  \bibinfo{pages}{9396} (\bibinfo{year}{2001}).

\bibitem[{\citenamefont{Chenoweth et~al.}(2008)\citenamefont{Chenoweth, van
  Duin, and Goddard}}]{Chenoweth2008reaxff}
\bibinfo{author}{\bibfnamefont{K.}~\bibnamefont{Chenoweth}},
  \bibinfo{author}{\bibfnamefont{A.~C.~T.} \bibnamefont{van Duin}},
  \bibnamefont{and} \bibinfo{author}{\bibfnamefont{W.~A.}
  \bibnamefont{Goddard}}, \bibinfo{journal}{J. Phys. Chem. A}
  \textbf{\bibinfo{volume}{112}}, \bibinfo{pages}{1040} (\bibinfo{year}{2008}).

\bibitem[{\citenamefont{Shan et~al.}(2010)\citenamefont{Shan, Devine, Kemper,
  Sinnott, and Phillpot}}]{Shan2010comb}
\bibinfo{author}{\bibfnamefont{T.-R.} \bibnamefont{Shan}},
  \bibinfo{author}{\bibfnamefont{B.~D.} \bibnamefont{Devine}},
  \bibinfo{author}{\bibfnamefont{T.~W.} \bibnamefont{Kemper}},
  \bibinfo{author}{\bibfnamefont{S.~B.} \bibnamefont{Sinnott}},
  \bibnamefont{and} \bibinfo{author}{\bibfnamefont{S.~R.}
  \bibnamefont{Phillpot}}, \bibinfo{journal}{Phys. Rev. B}
  \textbf{\bibinfo{volume}{81}}, \bibinfo{pages}{125328}
  (\bibinfo{year}{2010}).

\bibitem[{\citenamefont{Devine et~al.}(2011)\citenamefont{Devine, Shan, Cheng,
  McGaughey, Lee, Phillpot, Sinnott et~al.}}]{Devine2011comb}
\bibinfo{author}{\bibfnamefont{B.}~\bibnamefont{Devine}},
  \bibinfo{author}{\bibfnamefont{T.-R.} \bibnamefont{Shan}},
  \bibinfo{author}{\bibfnamefont{Y.-T.} \bibnamefont{Cheng}},
  \bibinfo{author}{\bibfnamefont{A.~J.} \bibnamefont{McGaughey}},
  \bibinfo{author}{\bibfnamefont{M.}~\bibnamefont{Lee}},
  \bibinfo{author}{\bibfnamefont{S.~R.} \bibnamefont{Phillpot}},
  \bibinfo{author}{\bibfnamefont{S.~B.} \bibnamefont{Sinnott}},
  \bibnamefont{et~al.}, \bibinfo{journal}{Phys. Rev. B}
  \textbf{\bibinfo{volume}{84}}, \bibinfo{pages}{125308}
  (\bibinfo{year}{2011}).

\bibitem[{\citenamefont{Powell}(1965)}]{Powell1965}
\bibinfo{author}{\bibfnamefont{M.~J.~D.} \bibnamefont{Powell}},
  \bibinfo{journal}{Comp. J.} \textbf{\bibinfo{volume}{7}},
  \bibinfo{pages}{303} (\bibinfo{year}{1965}).

\bibitem[{\citenamefont{Nelder and Mead}(1965)}]{Nelder1965}
\bibinfo{author}{\bibfnamefont{J.~A.} \bibnamefont{Nelder}} \bibnamefont{and}
  \bibinfo{author}{\bibfnamefont{R.}~\bibnamefont{Mead}},
  \bibinfo{journal}{Comput. J.} \textbf{\bibinfo{volume}{7}},
  \bibinfo{pages}{308} (\bibinfo{year}{1965}),
  \eprint{http://comjnl.oxfordjournals.org/content/7/4/308.full.pdf+html},
  \urlprefix\url{http://comjnl.oxfordjournals.org/content/7/4/308.abstract}.

\bibitem[{\citenamefont{Broyden}(1970)}]{Broyden1970}
\bibinfo{author}{\bibfnamefont{C.~G.} \bibnamefont{Broyden}},
  \bibinfo{journal}{IMA J. Appl. Math.} \textbf{\bibinfo{volume}{6}},
  \bibinfo{pages}{76} (\bibinfo{year}{1970}).

\bibitem[{\citenamefont{Fletcher}(1970)}]{Fletcher1970}
\bibinfo{author}{\bibfnamefont{R.}~\bibnamefont{Fletcher}},
  \bibinfo{journal}{Comput. J.} \textbf{\bibinfo{volume}{13}},
  \bibinfo{pages}{317} (\bibinfo{year}{1970}).

\bibitem[{\citenamefont{Goldfarb}(1970)}]{Goldfarb1970}
\bibinfo{author}{\bibfnamefont{D.}~\bibnamefont{Goldfarb}},
  \bibinfo{journal}{Math. Comput.} \textbf{\bibinfo{volume}{24}},
  \bibinfo{pages}{23} (\bibinfo{year}{1970}).

\bibitem[{\citenamefont{Shanno}(1970)}]{Shanno1970}
\bibinfo{author}{\bibfnamefont{D.~F.} \bibnamefont{Shanno}},
  \bibinfo{journal}{Math. Comput.} \textbf{\bibinfo{volume}{24}},
  \bibinfo{pages}{647} (\bibinfo{year}{1970}).

\bibitem[{\citenamefont{Mor{\'e}}(1978)}]{More1978LM}
\bibinfo{author}{\bibfnamefont{J.~J.} \bibnamefont{Mor{\'e}}}, in
  \emph{\bibinfo{booktitle}{Numerical analysis}}
  (\bibinfo{publisher}{Springer}, \bibinfo{year}{1978}), pp.
  \bibinfo{pages}{105--116}.

\bibitem[{\citenamefont{van Duin et~al.}(1994)\citenamefont{van Duin, Baas, and
  van~de Graaf}}]{vanDuin1994}
\bibinfo{author}{\bibfnamefont{A.~C.~T.} \bibnamefont{van Duin}},
  \bibinfo{author}{\bibfnamefont{J.~M.~A.} \bibnamefont{Baas}},
  \bibnamefont{and} \bibinfo{author}{\bibfnamefont{B.}~\bibnamefont{van~de
  Graaf}}, \bibinfo{journal}{J. Chem. Soc., Faraday Trans.}
  \textbf{\bibinfo{volume}{90}}, \bibinfo{pages}{2881} (\bibinfo{year}{1994}).

\bibitem[{\citenamefont{Diwekar}(2008)}]{Diwekar2008opt}
\bibinfo{author}{\bibfnamefont{U.}~\bibnamefont{Diwekar}},
  \emph{\bibinfo{title}{Introduction to applied optimization}}
  (\bibinfo{publisher}{Springer Science \& Business Media},
  \bibinfo{year}{2008}).

\bibitem[{\citenamefont{K{\i}nac{\i} et~al.}(2012)\citenamefont{K{\i}nac{\i},
  Haskins, Sevik, and {\c{C}}a{\u{g}}{\i}n}}]{Kinaci2012}
\bibinfo{author}{\bibfnamefont{A.}~\bibnamefont{K{\i}nac{\i}}},
  \bibinfo{author}{\bibfnamefont{J.~B.} \bibnamefont{Haskins}},
  \bibinfo{author}{\bibfnamefont{C.}~\bibnamefont{Sevik}}, \bibnamefont{and}
  \bibinfo{author}{\bibfnamefont{T.}~\bibnamefont{{\c{C}}a{\u{g}}{\i}n}},
  \bibinfo{journal}{Phys. Rev. B} \textbf{\bibinfo{volume}{86}},
  \bibinfo{pages}{115410} (\bibinfo{year}{2012}).

\bibitem[{\citenamefont{Saito et~al.}(2001)\citenamefont{Saito, Sasaki, Moriya,
  Kagatsume, and Noro}}]{Saito2001}
\bibinfo{author}{\bibfnamefont{Y.}~\bibnamefont{Saito}},
  \bibinfo{author}{\bibfnamefont{N.}~\bibnamefont{Sasaki}},
  \bibinfo{author}{\bibfnamefont{H.}~\bibnamefont{Moriya}},
  \bibinfo{author}{\bibfnamefont{A.}~\bibnamefont{Kagatsume}},
  \bibnamefont{and} \bibinfo{author}{\bibfnamefont{S.}~\bibnamefont{Noro}},
  \bibinfo{journal}{JSME Int. J. Series A Solid Mechanics and Material
  Engineering} \textbf{\bibinfo{volume}{44}}, \bibinfo{pages}{207}
  (\bibinfo{year}{2001}).

\bibitem[{\citenamefont{Narayanan et~al.}(2015)\citenamefont{Narayanan, Kinaci,
  Sen, Davis, Gray, Chan, and Sankaranarayanan}}]{Narayanan2015au}
\bibinfo{author}{\bibfnamefont{B.}~\bibnamefont{Narayanan}},
  \bibinfo{author}{\bibfnamefont{A.}~\bibnamefont{Kinaci}},
  \bibinfo{author}{\bibfnamefont{F.~G.} \bibnamefont{Sen}},
  \bibinfo{author}{\bibfnamefont{M.~J.} \bibnamefont{Davis}},
  \bibinfo{author}{\bibfnamefont{S.~K.} \bibnamefont{Gray}},
  \bibinfo{author}{\bibfnamefont{M.~K.} \bibnamefont{Chan}}, \bibnamefont{and}
  \bibinfo{author}{\bibfnamefont{S.~K.} \bibnamefont{Sankaranarayanan}},
  \bibinfo{journal}{J. Phys. Chem. C}  (\bibinfo{year}{2015}).

\bibitem[{\citenamefont{Cherukara et~al.}(2016)\citenamefont{Cherukara,
  Narayanan, Kinaci, Kiran, Gray, Chan, and Sankaranarayanan}}]{Cherukara2016}
\bibinfo{author}{\bibfnamefont{M.~J.} \bibnamefont{Cherukara}},
  \bibinfo{author}{\bibfnamefont{B.}~\bibnamefont{Narayanan}},
  \bibinfo{author}{\bibfnamefont{A.}~\bibnamefont{Kinaci}},
  \bibinfo{author}{\bibfnamefont{S.}~\bibnamefont{Kiran}},
  \bibinfo{author}{\bibfnamefont{S.~K.} \bibnamefont{Gray}},
  \bibinfo{author}{\bibfnamefont{M.~K.} \bibnamefont{Chan}}, \bibnamefont{and}
  \bibinfo{author}{\bibfnamefont{S.~K.} \bibnamefont{Sankaranarayanan}},
  \bibinfo{journal}{J. Phys. Chem. Lett.}  (\bibinfo{year}{2016}).

\bibitem[{\citenamefont{Sen et~al.}(2015)\citenamefont{Sen, Kinaci, Narayanan,
  Gray, Davis, Sankaranarayanan, and Chan}}]{Sen15}
\bibinfo{author}{\bibfnamefont{F.~G.} \bibnamefont{Sen}},
  \bibinfo{author}{\bibfnamefont{A.}~\bibnamefont{Kinaci}},
  \bibinfo{author}{\bibfnamefont{B.}~\bibnamefont{Narayanan}},
  \bibinfo{author}{\bibfnamefont{S.~K.} \bibnamefont{Gray}},
  \bibinfo{author}{\bibfnamefont{M.~J.} \bibnamefont{Davis}},
  \bibinfo{author}{\bibfnamefont{S.~K. R.~S.} \bibnamefont{Sankaranarayanan}},
  \bibnamefont{and} \bibinfo{author}{\bibfnamefont{M.~K.~Y.}
  \bibnamefont{Chan}}, \bibinfo{journal}{J. Mater. Chem. A}
  \textbf{\bibinfo{volume}{3}}, \bibinfo{pages}{18970} (\bibinfo{year}{2015}).

\bibitem[{\citenamefont{Cui et~al.}(2012)\citenamefont{Cui, Gao, Cui, and
  Qu}}]{Cui2012pso}
\bibinfo{author}{\bibfnamefont{Z.}~\bibnamefont{Cui}},
  \bibinfo{author}{\bibfnamefont{F.}~\bibnamefont{Gao}},
  \bibinfo{author}{\bibfnamefont{Z.}~\bibnamefont{Cui}}, \bibnamefont{and}
  \bibinfo{author}{\bibfnamefont{J.}~\bibnamefont{Qu}},
  \bibinfo{journal}{Modelling Simul. Mater. Sci. Eng.}
  \textbf{\bibinfo{volume}{20}}, \bibinfo{pages}{015014}
  (\bibinfo{year}{2012}).

\bibitem[{\citenamefont{Kandemir et~al.}(2016)\citenamefont{Kandemir,
  Yapicioglu, Kinaci, {\c{C}}a{\u{g}}{\i}n, and Sevik}}]{Kandemir2016pso}
\bibinfo{author}{\bibfnamefont{A.}~\bibnamefont{Kandemir}},
  \bibinfo{author}{\bibfnamefont{H.}~\bibnamefont{Yapicioglu}},
  \bibinfo{author}{\bibfnamefont{A.}~\bibnamefont{Kinaci}},
  \bibinfo{author}{\bibfnamefont{T.}~\bibnamefont{{\c{C}}a{\u{g}}{\i}n}},
  \bibnamefont{and} \bibinfo{author}{\bibfnamefont{C.}~\bibnamefont{Sevik}},
  \bibinfo{journal}{Nanotechnology} \textbf{\bibinfo{volume}{27}},
  \bibinfo{pages}{055703} (\bibinfo{year}{2016}).

\bibitem[{\citenamefont{Hobday et~al.}(1999)\citenamefont{Hobday, Smith, and
  BelBruno}}]{Hobday1999}
\bibinfo{author}{\bibfnamefont{S.}~\bibnamefont{Hobday}},
  \bibinfo{author}{\bibfnamefont{R.}~\bibnamefont{Smith}}, \bibnamefont{and}
  \bibinfo{author}{\bibfnamefont{J.}~\bibnamefont{BelBruno}},
  \bibinfo{journal}{Nucl. Inst. Methods Phys. Res. Sec. B}
  \textbf{\bibinfo{volume}{153}}, \bibinfo{pages}{247} (\bibinfo{year}{1999}).

\bibitem[{\citenamefont{Zhang and Trinkle}(2015)}]{Zhang2015}
\bibinfo{author}{\bibfnamefont{P.}~\bibnamefont{Zhang}} \bibnamefont{and}
  \bibinfo{author}{\bibfnamefont{D.~R.} \bibnamefont{Trinkle}},
  \bibinfo{journal}{Modelling Simul. Mater. Sci. Eng.}
  \textbf{\bibinfo{volume}{23}}, \bibinfo{pages}{065011}
  (\bibinfo{year}{2015}).

\bibitem[{\citenamefont{Narayanan et~al.}(2016)\citenamefont{Narayanan,
  Sasikumar, Mei, Kinaci, Sen, Davis, Gray, Chan, and
  Sankaranarayanan}}]{Narayanan2016zrn}
\bibinfo{author}{\bibfnamefont{B.}~\bibnamefont{Narayanan}},
  \bibinfo{author}{\bibfnamefont{K.}~\bibnamefont{Sasikumar}},
  \bibinfo{author}{\bibfnamefont{Z.-G.} \bibnamefont{Mei}},
  \bibinfo{author}{\bibfnamefont{A.}~\bibnamefont{Kinaci}},
  \bibinfo{author}{\bibfnamefont{F.~G.} \bibnamefont{Sen}},
  \bibinfo{author}{\bibfnamefont{M.~J.} \bibnamefont{Davis}},
  \bibinfo{author}{\bibfnamefont{S.~K.} \bibnamefont{Gray}},
  \bibinfo{author}{\bibfnamefont{M.~K.} \bibnamefont{Chan}}, \bibnamefont{and}
  \bibinfo{author}{\bibfnamefont{S.~K.} \bibnamefont{Sankaranarayanan}},
  \bibinfo{journal}{J. Phys. Chem. C} \textbf{\bibinfo{volume}{120}},
  \bibinfo{pages}{17475} (\bibinfo{year}{2016}).

\bibitem[{\citenamefont{Comninellis and Vercesi}(1991)}]{Comninellis1991}
\bibinfo{author}{\bibfnamefont{C.}~\bibnamefont{Comninellis}} \bibnamefont{and}
  \bibinfo{author}{\bibfnamefont{G.}~\bibnamefont{Vercesi}},
  \bibinfo{journal}{J. Appl. Electrochem.} \textbf{\bibinfo{volume}{21}},
  \bibinfo{pages}{335} (\bibinfo{year}{1991}).

\bibitem[{\citenamefont{Comninellis and Vercesi}(1979)}]{Beni1979}
\bibinfo{author}{\bibfnamefont{C.}~\bibnamefont{Comninellis}} \bibnamefont{and}
  \bibinfo{author}{\bibfnamefont{G.}~\bibnamefont{Vercesi}},
  \bibinfo{journal}{Nature} \textbf{\bibinfo{volume}{282}},
  \bibinfo{pages}{281} (\bibinfo{year}{1979}).

\bibitem[{\citenamefont{Song et~al.}(2008)\citenamefont{Song, Zhang, Ma, Shao,
  Baker, and Yi}}]{Song2008}
\bibinfo{author}{\bibfnamefont{S.}~\bibnamefont{Song}},
  \bibinfo{author}{\bibfnamefont{H.}~\bibnamefont{Zhang}},
  \bibinfo{author}{\bibfnamefont{X.}~\bibnamefont{Ma}},
  \bibinfo{author}{\bibfnamefont{Z.}~\bibnamefont{Shao}},
  \bibinfo{author}{\bibfnamefont{R.~T.} \bibnamefont{Baker}}, \bibnamefont{and}
  \bibinfo{author}{\bibfnamefont{B.}~\bibnamefont{Yi}}, \bibinfo{journal}{Int.
  J. Hydrogen Energy} \textbf{\bibinfo{volume}{33}}, \bibinfo{pages}{4955}
  (\bibinfo{year}{2008}).

\bibitem[{\citenamefont{Suntivich et~al.}(2011)\citenamefont{Suntivich, May,
  Gasteiger, Goodenough, and Shao-Horn}}]{Suntivich2011}
\bibinfo{author}{\bibfnamefont{J.}~\bibnamefont{Suntivich}},
  \bibinfo{author}{\bibfnamefont{K.~J.} \bibnamefont{May}},
  \bibinfo{author}{\bibfnamefont{H.~A.} \bibnamefont{Gasteiger}},
  \bibinfo{author}{\bibfnamefont{J.~B.} \bibnamefont{Goodenough}},
  \bibnamefont{and}
  \bibinfo{author}{\bibfnamefont{Y.}~\bibnamefont{Shao-Horn}},
  \bibinfo{journal}{Science} \textbf{\bibinfo{volume}{334}},
  \bibinfo{pages}{1383} (\bibinfo{year}{2011}).

\bibitem[{\citenamefont{Lee et~al.}(2012)\citenamefont{Lee, Suntivich, May,
  Perry, and Shao-Horn}}]{lee2012synthesis}
\bibinfo{author}{\bibfnamefont{Y.}~\bibnamefont{Lee}},
  \bibinfo{author}{\bibfnamefont{J.}~\bibnamefont{Suntivich}},
  \bibinfo{author}{\bibfnamefont{K.~J.} \bibnamefont{May}},
  \bibinfo{author}{\bibfnamefont{E.~E.} \bibnamefont{Perry}}, \bibnamefont{and}
  \bibinfo{author}{\bibfnamefont{Y.}~\bibnamefont{Shao-Horn}},
  \bibinfo{journal}{J. Phys. Chem. Lett.} \textbf{\bibinfo{volume}{3}},
  \bibinfo{pages}{399} (\bibinfo{year}{2012}).

\bibitem[{\citenamefont{Blochl}(1994)}]{Blochl94}
\bibinfo{author}{\bibfnamefont{P.~E.} \bibnamefont{Blochl}},
  \bibinfo{journal}{Phys. Rev. B} \textbf{\bibinfo{volume}{50}},
  \bibinfo{pages}{17953} (\bibinfo{year}{1994}), ISSN
  \bibinfo{issn}{1098-0121}.

\bibitem[{\citenamefont{Kresse and Hafner}(1994)}]{Kresse94}
\bibinfo{author}{\bibfnamefont{G.}~\bibnamefont{Kresse}} \bibnamefont{and}
  \bibinfo{author}{\bibfnamefont{J.}~\bibnamefont{Hafner}},
  \bibinfo{journal}{Phys. Rev. B} \textbf{\bibinfo{volume}{49}},
  \bibinfo{pages}{14251} (\bibinfo{year}{1994}), ISSN
  \bibinfo{issn}{0163-1829}.

\bibitem[{\citenamefont{Kresse and Furthmuller}(1996)}]{Kresse96}
\bibinfo{author}{\bibfnamefont{G.}~\bibnamefont{Kresse}} \bibnamefont{and}
  \bibinfo{author}{\bibfnamefont{J.}~\bibnamefont{Furthmuller}},
  \bibinfo{journal}{Comp. Mater. Sci.} \textbf{\bibinfo{volume}{6}},
  \bibinfo{pages}{15} (\bibinfo{year}{1996}), ISSN \bibinfo{issn}{0927-0256}.

\bibitem[{\citenamefont{Perdew et~al.}(1996)\citenamefont{Perdew, Burke, and
  Ernzerhof}}]{Perdew96}
\bibinfo{author}{\bibfnamefont{J.~P.} \bibnamefont{Perdew}},
  \bibinfo{author}{\bibfnamefont{K.}~\bibnamefont{Burke}}, \bibnamefont{and}
  \bibinfo{author}{\bibfnamefont{M.}~\bibnamefont{Ernzerhof}},
  \bibinfo{journal}{Phys. Rev. Lett.} \textbf{\bibinfo{volume}{77}},
  \bibinfo{pages}{3865} (\bibinfo{year}{1996}), ISSN \bibinfo{issn}{0031-9007}.

\bibitem[{\citenamefont{Kresse and Joubert}(1999)}]{Kresse99}
\bibinfo{author}{\bibfnamefont{G.}~\bibnamefont{Kresse}} \bibnamefont{and}
  \bibinfo{author}{\bibfnamefont{D.}~\bibnamefont{Joubert}},
  \bibinfo{journal}{Phys. Rev. B} \textbf{\bibinfo{volume}{59}},
  \bibinfo{pages}{1758} (\bibinfo{year}{1999}), ISSN \bibinfo{issn}{1098-0121}.

\bibitem[{\citenamefont{Liechtenstein et~al.}(1995)\citenamefont{Liechtenstein,
  Anisimov, and Zaanen}}]{Liechtenstein95}
\bibinfo{author}{\bibfnamefont{A.~I.} \bibnamefont{Liechtenstein}},
  \bibinfo{author}{\bibfnamefont{V.~I.} \bibnamefont{Anisimov}},
  \bibnamefont{and} \bibinfo{author}{\bibfnamefont{J.}~\bibnamefont{Zaanen}},
  \bibinfo{journal}{Phys. Rev. B} \textbf{\bibinfo{volume}{52}},
  \bibinfo{pages}{R5467} (\bibinfo{year}{1995}), ISSN
  \bibinfo{issn}{0163-1829}.

\bibitem[{\citenamefont{Dudarev et~al.}(1998)\citenamefont{Dudarev, Botton,
  Savrasov, Humphreys, and Sutton}}]{Dudarev98}
\bibinfo{author}{\bibfnamefont{S.~L.} \bibnamefont{Dudarev}},
  \bibinfo{author}{\bibfnamefont{G.~A.} \bibnamefont{Botton}},
  \bibinfo{author}{\bibfnamefont{S.~Y.} \bibnamefont{Savrasov}},
  \bibinfo{author}{\bibfnamefont{C.~J.} \bibnamefont{Humphreys}},
  \bibnamefont{and} \bibinfo{author}{\bibfnamefont{A.~P.}
  \bibnamefont{Sutton}}, \bibinfo{journal}{Phys. Rev. B}
  \textbf{\bibinfo{volume}{57}}, \bibinfo{pages}{1505} (\bibinfo{year}{1998}),
  ISSN \bibinfo{issn}{1098-0121}, \bibinfo{note}{times Cited: 2304 2317}.

\bibitem[{\citenamefont{Fujiwara et~al.}(2013)\citenamefont{Fujiwara, Fukuma,
  Matsuno, Idzuchi, Niimi, Otani, and Takagi}}]{Fujiwara13}
\bibinfo{author}{\bibfnamefont{K.}~\bibnamefont{Fujiwara}},
  \bibinfo{author}{\bibfnamefont{Y.}~\bibnamefont{Fukuma}},
  \bibinfo{author}{\bibfnamefont{J.}~\bibnamefont{Matsuno}},
  \bibinfo{author}{\bibfnamefont{H.}~\bibnamefont{Idzuchi}},
  \bibinfo{author}{\bibfnamefont{Y.}~\bibnamefont{Niimi}},
  \bibinfo{author}{\bibfnamefont{Y.}~\bibnamefont{Otani}}, \bibnamefont{and}
  \bibinfo{author}{\bibfnamefont{H.}~\bibnamefont{Takagi}},
  \bibinfo{journal}{Nat. Comm.} \textbf{\bibinfo{volume}{4}}
  (\bibinfo{year}{2013}), ISSN \bibinfo{issn}{2041-1723}.

\bibitem[{\citenamefont{Panda et~al.}(2014)\citenamefont{Panda, Bhowal, Delin,
  Eriksson, and Dasgupta}}]{Panda14}
\bibinfo{author}{\bibfnamefont{S.~K.} \bibnamefont{Panda}},
  \bibinfo{author}{\bibfnamefont{S.}~\bibnamefont{Bhowal}},
  \bibinfo{author}{\bibfnamefont{A.}~\bibnamefont{Delin}},
  \bibinfo{author}{\bibfnamefont{O.}~\bibnamefont{Eriksson}}, \bibnamefont{and}
  \bibinfo{author}{\bibfnamefont{I.}~\bibnamefont{Dasgupta}},
  \bibinfo{journal}{Phys. Rev. B} \textbf{\bibinfo{volume}{89}},
  \bibinfo{pages}{155102} (\bibinfo{year}{2014}).

\bibitem[{\citenamefont{Jain et~al.}(2013)\citenamefont{Jain, Ong, Hautier,
  Chen, Richards, Dacek, Cholia, Gunter, Skinner, Ceder
  et~al.}}]{Jain13MatProject}
\bibinfo{author}{\bibfnamefont{A.}~\bibnamefont{Jain}},
  \bibinfo{author}{\bibfnamefont{S.~P.} \bibnamefont{Ong}},
  \bibinfo{author}{\bibfnamefont{G.}~\bibnamefont{Hautier}},
  \bibinfo{author}{\bibfnamefont{W.}~\bibnamefont{Chen}},
  \bibinfo{author}{\bibfnamefont{W.~D.} \bibnamefont{Richards}},
  \bibinfo{author}{\bibfnamefont{S.}~\bibnamefont{Dacek}},
  \bibinfo{author}{\bibfnamefont{S.}~\bibnamefont{Cholia}},
  \bibinfo{author}{\bibfnamefont{D.}~\bibnamefont{Gunter}},
  \bibinfo{author}{\bibfnamefont{D.}~\bibnamefont{Skinner}},
  \bibinfo{author}{\bibfnamefont{G.}~\bibnamefont{Ceder}},
  \bibnamefont{et~al.}, \bibinfo{journal}{APL Materials}
  \textbf{\bibinfo{volume}{1}}, \bibinfo{pages}{011002} (\bibinfo{year}{2013}).

\bibitem[{\citenamefont{Mattsson et~al.}(2008)\citenamefont{Mattsson, Armiento,
  Paier, Kresse, Wills, and Mattsson}}]{Mattsson08AM05}
\bibinfo{author}{\bibfnamefont{A.~E.} \bibnamefont{Mattsson}},
  \bibinfo{author}{\bibfnamefont{R.}~\bibnamefont{Armiento}},
  \bibinfo{author}{\bibfnamefont{J.}~\bibnamefont{Paier}},
  \bibinfo{author}{\bibfnamefont{G.}~\bibnamefont{Kresse}},
  \bibinfo{author}{\bibfnamefont{J.~M.} \bibnamefont{Wills}}, \bibnamefont{and}
  \bibinfo{author}{\bibfnamefont{T.~R.} \bibnamefont{Mattsson}},
  \bibinfo{journal}{J. Chem. Phys.} \textbf{\bibinfo{volume}{128}},
  \bibinfo{pages}{084714} (\bibinfo{year}{2008}).

\bibitem[{\citenamefont{Heyd and Scuseria}(2004)}]{Heyd04HSE}
\bibinfo{author}{\bibfnamefont{J.}~\bibnamefont{Heyd}} \bibnamefont{and}
  \bibinfo{author}{\bibfnamefont{G.~E.} \bibnamefont{Scuseria}},
  \bibinfo{journal}{J. Chem. Phys.} \textbf{\bibinfo{volume}{121}},
  \bibinfo{pages}{1187} (\bibinfo{year}{2004}).

\bibitem[{\citenamefont{Press}(2007)}]{Press2007}
\bibinfo{author}{\bibfnamefont{W.~H.} \bibnamefont{Press}},
  \emph{\bibinfo{title}{Numerical recipes 3rd edition: The art of scientific
  computing}} (\bibinfo{publisher}{Cambridge university press},
  \bibinfo{year}{2007}).

\bibitem[{\citenamefont{Huang et~al.}(1998)\citenamefont{Huang, Aine, Supek,
  Best, Ranken, and Flynn}}]{Huang1998Simplex}
\bibinfo{author}{\bibfnamefont{M.}~\bibnamefont{Huang}},
  \bibinfo{author}{\bibfnamefont{C.~J.} \bibnamefont{Aine}},
  \bibinfo{author}{\bibfnamefont{S.}~\bibnamefont{Supek}},
  \bibinfo{author}{\bibfnamefont{E.}~\bibnamefont{Best}},
  \bibinfo{author}{\bibfnamefont{D.}~\bibnamefont{Ranken}}, \bibnamefont{and}
  \bibinfo{author}{\bibfnamefont{E.}~\bibnamefont{Flynn}},
  \bibinfo{journal}{Electroencephalography and Clinical Neurophysiology/Evoked
  Potentials Section} \textbf{\bibinfo{volume}{108}}, \bibinfo{pages}{32}
  (\bibinfo{year}{1998}).

\bibitem[{\citenamefont{Mor{\'e} et~al.}(1980)\citenamefont{Mor{\'e}, Garbow,
  and Hillstrom}}]{More1980minpack}
\bibinfo{author}{\bibfnamefont{J.~J.} \bibnamefont{Mor{\'e}}},
  \bibinfo{author}{\bibfnamefont{B.~S.} \bibnamefont{Garbow}},
  \bibnamefont{and} \bibinfo{author}{\bibfnamefont{K.~E.}
  \bibnamefont{Hillstrom}}, \bibinfo{type}{Tech. Rep.},
  \bibinfo{institution}{Argonne National Laboratory} (\bibinfo{year}{1980}).

\bibitem[{\citenamefont{Mor{\'e} et~al.}(1984)\citenamefont{Mor{\'e}, Sorensen,
  Hillstrom, and Garbow}}]{More1984minpack}
\bibinfo{author}{\bibfnamefont{J.~J.} \bibnamefont{Mor{\'e}}},
  \bibinfo{author}{\bibfnamefont{D.~C.} \bibnamefont{Sorensen}},
  \bibinfo{author}{\bibfnamefont{K.~E.} \bibnamefont{Hillstrom}},
  \bibnamefont{and} \bibinfo{author}{\bibfnamefont{B.~S.}
  \bibnamefont{Garbow}}, \bibinfo{journal}{Sources and Development of
  Mathematical Software} pp. \bibinfo{pages}{88--111} (\bibinfo{year}{1984}).

\bibitem[{\citenamefont{Wild}(2014)}]{SWCHAP14}
\bibinfo{author}{\bibfnamefont{S.~M.} \bibnamefont{Wild}},
  \bibinfo{type}{Preprint} \bibinfo{number}{ANL/MCS-P5120-0414},
  \bibinfo{institution}{Argonne National Laboratory, Mathematics and Computer
  Science Division} (\bibinfo{year}{2014}),
  \urlprefix\url{http://www.mcs.anl.gov/papers/P5120-0414.pdf}.

\bibitem[{\citenamefont{Wild et~al.}(2015)\citenamefont{Wild, Sarich, and
  Schunck}}]{WSS15}
\bibinfo{author}{\bibfnamefont{S.~M.} \bibnamefont{Wild}},
  \bibinfo{author}{\bibfnamefont{J.}~\bibnamefont{Sarich}}, \bibnamefont{and}
  \bibinfo{author}{\bibfnamefont{N.}~\bibnamefont{Schunck}},
  \bibinfo{journal}{Journal of Physics G: Nuclear and Particle Physics}
  \textbf{\bibinfo{volume}{42}}, \bibinfo{pages}{034031}
  (\bibinfo{year}{2015}).

\bibitem[{\citenamefont{PETSc}()}]{Petsc}
\bibinfo{author}{\bibnamefont{PETSc}}, \emph{\bibinfo{title}{Extensible toolkit
  for scientific computation}},
  \bibinfo{howpublished}{\url{http://mcs.anl.gov/petsc/}}.

\bibitem[{\citenamefont{Goldberg et~al.}(1989)\citenamefont{Goldberg, Korb, and
  Deb}}]{Goldberg1989}
\bibinfo{author}{\bibfnamefont{D.}~\bibnamefont{Goldberg}},
  \bibinfo{author}{\bibfnamefont{B.}~\bibnamefont{Korb}}, \bibnamefont{and}
  \bibinfo{author}{\bibfnamefont{K.}~\bibnamefont{Deb}},
  \bibinfo{journal}{Complex Systems} \textbf{\bibinfo{volume}{3}},
  \bibinfo{pages}{493} (\bibinfo{year}{1989}).

\bibitem[{\citenamefont{Sastry and Goldberg}(2001)}]{Sastry01}
\bibinfo{author}{\bibfnamefont{K.}~\bibnamefont{Sastry}} \bibnamefont{and}
  \bibinfo{author}{\bibfnamefont{D.}~\bibnamefont{Goldberg}},
  \bibinfo{journal}{Intelligent Engineering Systems Through Artificial Neural
  Networks} \textbf{\bibinfo{volume}{11}}, \bibinfo{pages}{129}
  (\bibinfo{year}{2001}).

\bibitem[{\citenamefont{Deb and Agrawal}(1995)}]{Deb95-crossover}
\bibinfo{author}{\bibfnamefont{K.}~\bibnamefont{Deb}} \bibnamefont{and}
  \bibinfo{author}{\bibfnamefont{R.~B.} \bibnamefont{Agrawal}},
  \bibinfo{journal}{Complex Systems} \textbf{\bibinfo{volume}{9}},
  \bibinfo{pages}{115} (\bibinfo{year}{1995}), ISSN \bibinfo{issn}{0891-2513},
  \urlprefix\url{<Go to ISI>://INSPEC:5209474}.

\bibitem[{\citenamefont{Deb and Kumar}(1995)}]{Deb95-crossover2}
\bibinfo{author}{\bibfnamefont{K.}~\bibnamefont{Deb}} \bibnamefont{and}
  \bibinfo{author}{\bibfnamefont{A.}~\bibnamefont{Kumar}},
  \bibinfo{journal}{Complex Systems} \textbf{\bibinfo{volume}{9}},
  \bibinfo{pages}{431} (\bibinfo{year}{1995}), ISSN \bibinfo{issn}{0891-2513}.

\bibitem[{\citenamefont{Sastry}(2007)}]{Sastry07GA}
\bibinfo{author}{\bibfnamefont{K.}~\bibnamefont{Sastry}}, \bibinfo{type}{Tech.
  Rep.}, \bibinfo{institution}{University of Illinois at Urbana-Champaign}
  (\bibinfo{year}{2007}).

\bibitem[{\citenamefont{Deb et~al.}(2002)\citenamefont{Deb, Pratap, Agrawal,
  and Meyarivan}}]{Deb02-nsga2}
\bibinfo{author}{\bibfnamefont{K.}~\bibnamefont{Deb}},
  \bibinfo{author}{\bibfnamefont{A.}~\bibnamefont{Pratap}},
  \bibinfo{author}{\bibfnamefont{S.}~\bibnamefont{Agrawal}}, \bibnamefont{and}
  \bibinfo{author}{\bibfnamefont{T.}~\bibnamefont{Meyarivan}},
  \bibinfo{journal}{Trans. Evol. Comp} \textbf{\bibinfo{volume}{6}},
  \bibinfo{pages}{182} (\bibinfo{year}{2002}).

\bibitem[{\citenamefont{Ehrgott}(2005)}]{Ehrgott2005opt}
\bibinfo{author}{\bibfnamefont{M.}~\bibnamefont{Ehrgott}},
  \emph{\bibinfo{title}{Multicriteria optimization}}
  (\bibinfo{publisher}{Springer Science \& Business Media},
  \bibinfo{year}{2005}).

\bibitem[{\citenamefont{Konak et~al.}(2006)\citenamefont{Konak, Coit, and
  Smith}}]{Konak2006multi}
\bibinfo{author}{\bibfnamefont{A.}~\bibnamefont{Konak}},
  \bibinfo{author}{\bibfnamefont{D.~W.} \bibnamefont{Coit}}, \bibnamefont{and}
  \bibinfo{author}{\bibfnamefont{A.~E.} \bibnamefont{Smith}},
  \bibinfo{journal}{Reliability Engineering \& System Safety}
  \textbf{\bibinfo{volume}{91}}, \bibinfo{pages}{992} (\bibinfo{year}{2006}).

\end{thebibliography}

\end{document}